\documentclass[aps, prd, twocolumn, superscriptaddress, preprintnumbers, nofootinbib, longbibliography]{revtex4-1}

\usepackage{amsmath}	
\usepackage{amsthm}		
\usepackage{amssymb}	
\usepackage{datetime}
\usepackage{graphicx}
\usepackage{color}
\usepackage{verbatim}
\usepackage{bm}

\usepackage{titlesec}
\titlespacing{\section}{0pt}{10ex}{2ex}

\usepackage[notrig]{physics}

\usepackage[colorlinks=true, citecolor=midblue, linkcolor=midblue, urlcolor=midblue]{hyperref}

\usepackage{setspace}

\usepackage[svgnames]{xcolor}

\usepackage{tikz,xcolor}
\definecolor{lime}{HTML}{A6CE39}
\DeclareRobustCommand{\orcidicon}{
	\begin{tikzpicture}
	\draw[lime, fill=lime] (0,0) 
	circle [radius=0.16] 
	node[white] {{\fontfamily{qag}\selectfont \tiny ID}};
	\draw[white, fill=white] (-0.0625,0.095) 
	circle [radius=0.007];
	\end{tikzpicture}
	\hspace{-2mm}
}
\foreach \x in {A, ..., Z}{
	\expandafter\xdef\csname orcid\x\endcsname{\noexpand\href{https://orcid.org/\csname orcidauthor\x\endcsname}{\noexpand\orcidicon}}
}

\definecolor{grey}{rgb}{0.4,0.4,0.4}
\definecolor{dullmagenta}{rgb}{0.4,0,0.4}
\definecolor{darkblue}{rgb}{0,0,0.4}
\definecolor{midblue}{rgb}{0,0,0.5}
\definecolor{midred}{rgb}{0.5,0,0}
\definecolor{orange}{rgb}{1,0.5,0}
\definecolor{lightbrown}{rgb}{0.75,0.5,0.25}
\definecolor{tan}{cmyk}{0.14,0.42,0.56,0}
\definecolor{djunglegreen}{cmyk}{0.99,0,0.52,0}
\definecolor{lightgreen}{rgb}{0,1,0}
\definecolor{olivegreen}{cmyk}{0.64,0,0.95,0.40}
\definecolor{midgreen}{rgb}{0.0,0.675,0.0}
\definecolor{darkgreen}{rgb}{0,0.5,0}

\settimeformat{ampmtime}

\settimeformat{ampmtime}

\usepackage[font=small,labelfont=bf,justification=raggedright]{caption}
\usepackage{subcaption}

\newcommand{\FirstAffiliation}{\affiliation{
	Arnold Sommerfeld Center,
	Ludwig-Maximilians-Universit{\"a}t,
	Theresienstra{\ss}e 37,
	80333 M{\"u}nchen,
	Germany}}
 
\newcommand{\SecondAffiliation}{\affiliation{
	Max-Planck-Institut f{\"u}r Physik,
	Boltzmannstra{\ss}e 8,
	85748 Garching,
	Germany}}

\newcommand{\ThirdAffiliation}{\affiliation{
    Department of Particle Physics and Astrophysics, Weizmann Institute of Science, Rehovot 7610001, Israel}}

\newcommand{\FourthAffiliation}{\affiliation{
Institut de F\'isica d’Altes Energies (IFAE) and The Barcelona Institute of Science and Technology (BIST),
Campus UAB, 08193 Bellaterra (Barcelona), Spain}}

\date{\formatdate{\day}{\month}{\year}, \currenttime}

\begin{document}

\title{Simulations of magnetic monopole collisions
}

\author{Maximilian Bachmaier\orcidB{}}
\email{maximilian.bachmaier@physik.uni-muenchen.de}
\FirstAffiliation
\SecondAffiliation

\author{Gia Dvali}
\FirstAffiliation
\SecondAffiliation

\author{Josef Seitz\orcidC{}}
\email{josef-emanuel.seitz@weizmann.ac.il}
\ThirdAffiliation

\author{Juan Sebasti\'an Valbuena-Berm\'udez\orcidA{}}
\email{jvalbuena@ifae.es}
\FourthAffiliation

\date{\small\today} 

\begin{abstract} 

\noindent In this paper, we investigate the scattering of BPS magnetic monopoles through numerical simulations. We present an ansatz for various multi-monopole configurations suitable for analyzing monopole scattering processes. Our study includes planar scattering scenarios involving two, three, and four monopoles, as well as non-planar processes where three and four monopoles form intermediate tetrahedral and cubic states, respectively. Our observations align with the theoretical predictions of the moduli space approximation. Furthermore, we extend our analysis to relativistic velocities and explore parameters beyond the BPS limit.

\end{abstract}
\keywords{Magnetic Monopoles, Moduli Approximation, Monopole Scattering}

\maketitle
    
\section{\MakeUppercase{Introduction}}
\label{sec:introduction}


Magnetic monopoles of 't Hooft and Polyakov~\cite{tHooft:1974kcl, Polyakov:1974ek} represent consistent non-singular 
solitonic realizations of the idea of an isolated magnetic pole 
by Dirac.  They appear as topologically stable solutions in gauge theories with non-trivial second homotopy group of the vacuum manifold $\pi_2(\mathcal{M})$.

Monopoles play a very important role in physics. 
In particular, they are an intrinsic part of all grand unified theories. 
Although the cosmological monopole abundance can be strongly suppressed either by the 
period of inflation~\cite{Guth:1980zm} or by a mechanism of non-restoration of grand unified symmetry at high temperature
\cite{Dvali:1995cj}, in many motivated inflationary scenarios the phase transition with 
grand unified symmetry breaking does take place after inflation \cite{Dvali:1994ms}. In such post-inflationary scenarios the formation of
monopoles is unavoidable and their cosmological impact is determined by their subsequent 
interactions with each other and with other topological defects \footnote{In particular, it has been argued that 
the monopole number can be reduced either by their annihilation \cite{Langacker:1980kd} or 
via the mechanism of so-called ``erasure" \cite{Dvali:1997sa, Dvali:2022rgx} 
in which monopole unwinds upon its encounter with a (collapsing) domain wall (for recent analysis, see \cite{Bachmaier:2023zmq} and references therein).}.   
 
It is also commonly accepted that monopoles play a crucial role in various confining phases of gauge theories.  Therefore, it is fundamental to better understand various regimes of monopole interactions. 
The purpose of the present work is to contribute to this effort.

One of the most significant developments in the study of monopoles is the exact solution provided by Prasad and Sommerfield~\cite{Prasad-Sommerfield:1975} in the Bogomolny-Prasad-Sommerfield (BPS) limit~\cite{Prasad-Sommerfield:1975, Bogomolny:1975de}. This solution exhibits a remarkable cancellation between the Coulomb and scalar forces of two magnetic monopoles of the same magnetic charge~\cite{Manton:1977er}, simplifying 
the analysis of the interactions of two or more monopoles. 

The low-energy dynamics of BPS magnetic monopoles, including multi-monopole configurations, can be described by the geodesic motion in the moduli space. This idea, introduced by Manton~\cite{Manton:1981mp} and later proved by Stuart~\cite{Stuart:1994tc}, provides a valuable tool for analyzing monopole interactions in the moduli space approximation.
For two monopoles, the moduli space is described by the Atiyah-Hitchin metric~\cite{Atiyah:1985dv, Atiyah-Hitchin1988}.

The detailed analysis of this metric revealed several scattering scenarios, some of which we will discuss in this paper using numerical simulations. In order to facilitate this study, we developed approximate analytic configurations that serve as initial conditions for our simulations.
We simulated the right-angle scattering of two BPS magnetic monopoles and extended our analysis to relativistic scattering speeds and parameters beyond the BPS limit.
Additionally, we examined the $60$-degree and $45$-degree scattering of three and four monopoles, respectively. Our investigation further enabled us to explore non-planar scattering scenarios involving toroidal $N$-monopoles interacting with either a charge-one monopole or another toroidal monopole.

The paper is structured as follows. First (sections \ref{sec:magnetic-monopole-solution}, \ref{sec:moduli-space-for-magnetic-monopoles}, \ref{sec:limits-of-the-moduli-approximation}), we are providing some theoretical background of 't Hooft-Polyakov magnetic monopoles and the moduli space approximation. In section~\ref{sec:charge-n-magnetic-monopoles}, we are numerically analyzing the toroidal structure of the charge-$N$ monopole. Subsequently, we are presenting approximate analytic configurations for two (section~\ref{sec:two-monopole-configuration}) and more than two (section~\ref{sec:n-monopole-configuration}) monopoles that can be used to analyze the scattering of monopoles within a plane. In section~\ref{sec:results}, we are giving the results of our numerical simulations of such scattering processes. Finally  (section~\ref{sec:non-planar-scattering}), we present the numerical results of some non-planar scattering cases.

\section{\MakeUppercase{Magnetic Monopole Solution}}
\label{sec:magnetic-monopole-solution}
The minimal model that contains 't Hooft–Polyakov magnetic monopoles is an $SU(2)$ gauge theory with a scalar field $\phi$ transforming in the adjoint representation~\cite{tHooft:1974kcl, Polyakov:1974ek}. The vacuum manifold is given by $\mathcal{M} = SU(2)/U(1) \cong S^2$, which possesses a non-trivial second homotopy group $\pi_2(\mathcal{M}) \cong \mathbb{Z}$, permitting magnetic monopole solutions.
The Lagrangian of the model is
\begin{align}
    \label{eq:Lagrangian}
  \mathcal{L}=-\frac{1}{2}\Tr\left(G_{\mu\nu}G^{\mu\nu} \right)+\Tr\left((D_\mu\phi)^\dagger (D^\mu\phi) \right)-V(\phi),
\end{align}
with the Higgs potential
\begin{align}
  \label{eq:potential}
  V(\phi)=\lambda\left(\Tr(\phi^\dagger\phi)-\frac{v^2}{2}\right)^2.
\end{align}
The scalar field can be expressed in the form of a 
Hermitian matrix $\phi = \phi^a T_a$, where the $SU(2)$ generators $T_a$ are normalized as  $\Tr(T_a T_b)=\frac{1}{2}\delta_{ab}$.
The field strength tensor is
\begin{align}
  G_{\mu\nu}\equiv\partial_{\mu}W_{\nu}-\partial_{\nu}W_{\mu}-ig \left[ W_\mu, W_\nu \right],
\end{align}
with the gauge fields $W_\mu= W^a_\mu T_a$.
The covariant derivative has the form 
\begin{align}
  D_{\mu}\phi\equiv\partial_{\mu}\phi-ig\left[ W_\mu,\phi \right].
\end{align}
From the Lagrangian, the following field equations can be derived
\begin{align}
\label{eq:field-equations}
  &D_{\mu}(D^\mu\phi)^a+ \pdv{V}{\phi^a}=0,\nonumber\\
  &D_{\mu}G^{a\mu\nu}-g\varepsilon_{abc}\ (D^\nu \phi)^b \phi^c=0.
\end{align}
The vacuum of the theory Higgses the gauge group $SU(2)$ down to the $U(1)$ subgroup. The resulting spectrum 
is the following. 
One real gauge field, corresponding to the unbroken $U(1)$, remains massless. The two other real components of the gauge field gain masses $m_v = v g$
and form a complex vector boson charged under $U(1)$. Additionally, there is a real Higgs field with mass $m_h = \sqrt{2\lambda} v$.

A scalar field configuration that describes a non-trivial mapping $S_2 \rightarrow S_2$ is given by
\begin{align}
\label{eq:scalar-field-monopole}
    \phi^a=\frac{1}{g}\frac{r^a}{r^2}H(r),
\end{align}
where $H(r)$ is some profile function with the boundary conditions $H(r)/m_v r \xrightarrow{r \rightarrow \infty} 1$ and $H(r)\xrightarrow{r \rightarrow 0}0$.
To ensure finite energy, one has to demand $(D_i \phi)^a\xrightarrow{r\rightarrow\infty}0$. This condition is satisfied by the following choice of the gauge fields
\begin{align}
\label{eq:gauge-field-monopole}
     W^a_i&=\frac{1}{g}\varepsilon_{aij}\frac{r^j}{r^2}(1-K(r)),\nonumber\\
    W^a_t&=0,
\end{align}
with $K(r)\xrightarrow{r \rightarrow \infty}0$ and $K(r)\xrightarrow{r \rightarrow 0}1$.
We can insert this ansatz into the gauge-invariant field strength tensor~\cite{tHooft:1974kcl}
\begin{align}
    F_{\mu\nu}=G_{\mu\nu}^a \hat{\phi}^a-\frac{1}{g}\varepsilon_{abc}\hat{\phi}^a (D_\mu \hat{\phi})^b (D_\nu \hat{\phi})^c,
\end{align}
and calculate the long-range magnetic field
\begin{align}
    B_i=-\frac{1}{2}\varepsilon_{ijk} F_{jk}\xrightarrow{r\rightarrow \infty}\frac{1}{g}\frac{x^i}{r^3}.
\end{align}
This magnetic field corresponds to a magnetic monopole of magnetic charge $\frac{4 \pi}{g}$. \\

By inserting the ansatz~\eqref{eq:scalar-field-monopole},~\eqref{eq:gauge-field-monopole} into the field equations~\eqref{eq:field-equations}, we 
obtain a system of differential equations for the profile functions
\begin{align}
  K''=&\frac{1}{r^2}\left(K^3-K+H^2 K\right),\nonumber\\
  H''=&\frac{2}{r^2}H K^2+\frac{m_h^2}{2} \left(\frac{H^2}{m_v^2 r^2}-1\right)H.
\end{align}
In general, these equations have no known analytical solution. However, Prasad and Sommerfield discovered a solution in the so-called BPS limit, in which  $m_h \to 0$ while $v$ is kept  fixed~\cite{Bogomolny:1975de, Prasad-Sommerfield:1975}.
The solution is given by
\begin{align}
\label{eq:BPS-solution}
    K(r)&=\frac{m_v r}{\sinh (m_v r)},\nonumber\\
    H(r)&=m_v r \coth (m_v r)-1.
\end{align}
A special property of the BPS limit is that the Higgs mass vanishes. This implies that the scalar interaction mediated by 
Higgs is long-range. In the BPS limit, the field equations are simplified to
\begin{align}
\label{eq:BPS-equation}
    B_i^a = (D_i \phi)^a,
\end{align}
indicating that the strength of the scalar interaction matches the strength of the magnetic force. Since the scalar interaction is always attractive, the magnetic force between two magnetic monopoles of the same charge is exactly compensated by the scalar force~\cite{Manton:1977er}.\\

\section{\MakeUppercase{Moduli Space for Magnetic Monopoles}}

\label{sec:moduli-space-for-magnetic-monopoles}

Magnetic monopole configurations are solitons, i.e. classical solutions of the field equations describing (several) localized lumps of energy. In the case of BPS configurations, they describe a minimal-energy solution within a given magnetic charge sector. In general, there will be many solutions with fixed values 
of total magnetic charge.  Let us focus on the solutions of equal energy that 
are obtained from one another by continuous deformations. 
The simplest example of such a deformation is a spatial translation of a given solution. We refer to the parameters describing such continuous deformations as \textit{collective coordinates} $X^{\alpha}$. An entire moduli space of solutions is swept out by the collective coordinates. 

For the case of magnetic monopoles of the $SU(2)$ theory at hand, there is an especially rich structure in the BPS limit $\lambda \rightarrow 0$. Notice that the BPS monopole equations are identical to the BPS equations of the $\mathcal{N}=2$ supersymmetric 
$SU(2)$ theory in four dimensions. 
The monopole configurations of our model are therefore also BPS in the supersymmetric sense.  Due to this connection,
we can mentally think of the above theory as a bosonic sector of its supersymmetric counterpart.  We can then use 
``spectator" supersymmetry for deriving the properties of the classical moduli space. 

Supersymmetry implies that the moduli space must be Hyperkaehler\footnote{Naively, from $\mathcal{N}=2$, one only gets a Kaehler property of the moduli space. However, the BPS equations are identical to the 4D Instanton equations and inherit a Hyperkaehler property from there \cite{Tong:2005un}.}, greatly restricting (and sometimes uniquely specifying) the moduli space in each charge sector. 

The moduli space structure leads to some surprising features such as a right-angle scattering of monopoles that we 
show numerically in the present work. We briefly review some aspects of 2-monopole spaces that are necessary for our discussion and refer to the reviews \cite{Manton:2004tk, Atiyah-Hitchin1988} as well as the lecture notes \cite{Tong:2005un} for more general expositions. 

After finding the solution \eqref{eq:BPS-solution}, one can discover `nearby' solutions by linearizing the BPS equation \eqref{eq:BPS-equation} around the background of the initial solution. In that way, one finds zero modes\footnote{The $W_0$-component is fixed by a gauge-fixing condition.} $\delta_{\alpha} \phi$,  $\delta_{\alpha} W_{i}$ that gives rise to a metric on the moduli space via computing their overlap:
\begin{equation}
    g_{\alpha \beta} = \text{Tr} \int d^3x (\delta_{\alpha} W_i\, \delta_{\beta} W_i + \delta_{\alpha} \phi \, \delta_{\beta} \phi).
\end{equation}
These zero modes are the infinitesimal counterpart of the collective coordinates.

One can now discuss the effective low energy physics utilizing the \text{moduli space approximation} \cite{Manton:1981mp} in which one assumes that the field configuration evolves along the moduli space. This evolution is accounted by the time-dependence of the collective coordinates:
\begin{equation}
    W_{\mu} = W_{\mu}(X^{\alpha}(t)), \quad \phi=\phi(X^{\alpha}(t)). 
\end{equation}
In this treatment, the action reduces to an action describing a point particle moving on the moduli space
\begin{align}
    S=\int \dd t \left(M+\frac{1}{2} g_{\alpha \beta}\dot{X}^\alpha \dot{X}^\beta \right). 
\end{align}
The low energy dynamics is thus described by a geodesic motion on the moduli space. 
The moduli space approximation clearly has a 
limited domain of validity; we refer to section \ref{sec:limits-of-the-moduli-approximation} for further discussions. 

 In this framework, one can study the low-energy scattering of monopoles by first finding the moduli space and the point on it which corresponds to two or more coincident monopoles, and then choosing an initial condition for a point particle on the moduli space such that it traverses through the coincidence point. 

As a warm-up exercise, we consider the moduli space of a single monopole. The three collective coordinates correspond to translations of the monopole core.\footnote{These can be thought of as the Goldstone modes of the space translation symmetries that are spontaneously broken by the monopole solution.}
But there needs to be at least one additional mode: the moduli space is Hyperkaehler, and as such, has a dimension with a multiple of four. Indeed, the fourth mode is given by `gauge transformations' \footnote{Since the transformation does not decay at infinity, `asymptotic symmetry' would be the more appropriate term.}
\begin{equation}\label{eq: electric charge transformation}
    U = \exp \left( \frac{i \phi \chi(t)}{v} \right). 
\end{equation}
These have an interesting physical interpretation: while such gauge transformation does not affect $\phi$, it has the effect that it induces an electric field $\sim \dot{\chi}$. Moving on the $U(1)$ swept out by $\chi$ thus corresponds to the monopole carrying electric charge. It turns out these are the only collective coordinates; the moduli space is given by
\begin{equation}
    \mathcal{M}= \mathbb{R}^3 \times S^1. 
\end{equation}

Now we turn to the case of two monopoles. There are again four obvious collective coordinates: three from the center of mass of the two-monopole configuration and one from the transformation \eqref{eq: electric charge transformation}. One expects that there should be more: at large distances, as there are no leading monopole-monopole interactions due to the BPS condition, there should be additional approximate collective coordinates describing the relative separation (and relative electric charge). Indeed, using the Hyperkaehler property and the asymptotic behaviour, one can show \cite{Atiyah-Hitchin1988,Manton:2004tk} that the nontrivial piece in the moduli space describing the relative dynamics of the monopoles is given by the Atiyah-Hitchin manifold:
\begin{equation}
    \mathcal{M} = \mathbb{R}^3 \times \frac{S^1 \times \mathcal{M}_{AH}}{\mathbb{Z}_2}.
\end{equation}
The Atiyah-Hitchin manifold is Hyperkaehler and respects the symmetries of the underlying problem (like an $SU(2)$ global symmetry inherited from the gauge group); we will not need its explicit form here.
An important geodesic submanifold of the Atiyah-Hitchin manifold is given by a surface ($\Sigma_1$ in the notation of \cite{Atiyah-Hitchin1988}) with the metric
\begin{equation}
    ds^2 = d\xi^2 + a^2 d\varphi^2,
\end{equation}
with $\varphi$ being in the coordinate range $[0,\pi]$, $a(\xi) = \xi$ asymptotically, and $a(\xi \rightarrow 0)$ going to zero faster than linear \cite{Atiyah-Hitchin1988}. 

$\Sigma_1$ describes a smoothed-out cone with (asymptotic) deficit angle $\pi$. In other words, asymptotically $\Sigma_1$ is described by $\mathbb{R}^2/\mathbb{Z}_2$. Scattering on that geodesic submanifold occurs when we restrict the motion of the monopoles onto a plane (denoted as $x$-$y$-plane from now on). 

\begin{figure}[h]
    \includegraphics[trim=60 20 60 120,clip,width=0.48\linewidth]{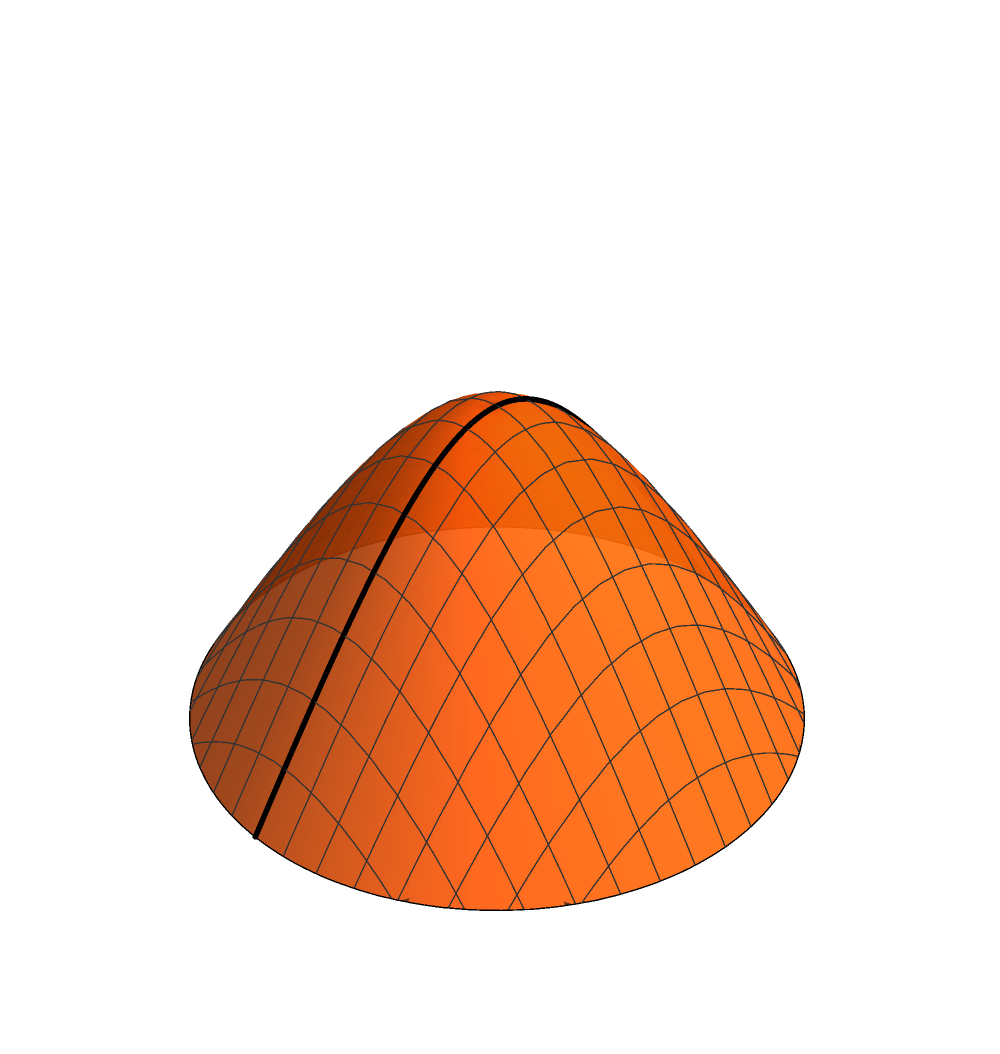}
    \includegraphics[trim=50 55 50 30,clip,width=0.48\linewidth]{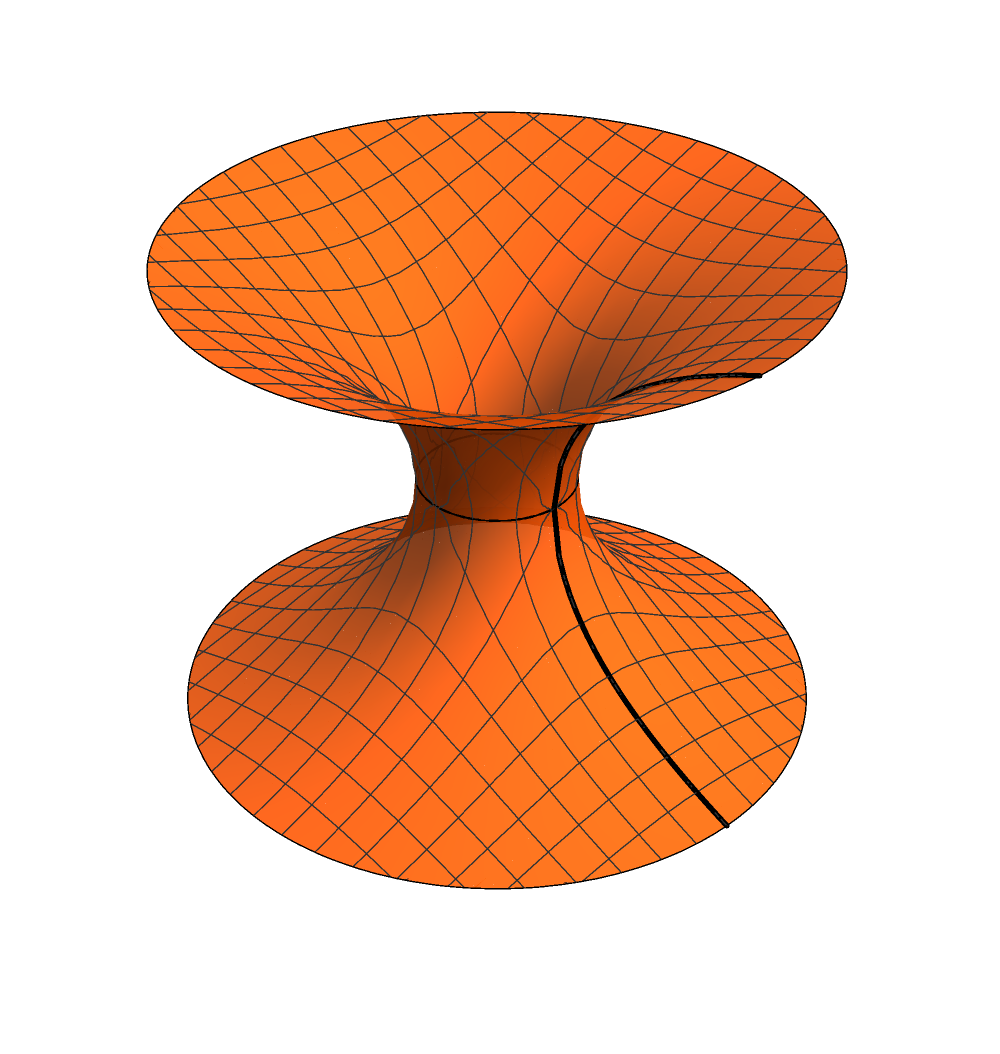}
    \caption{These two figures sketch the moduli space structure of two magnetic monopoles. The cone with the smoothed-out tip describes the motion of the monopoles in the $x$-$y$-plane. The trumped describes the motion in the other two planes.}
    \label{fig:moduli-space}
\end{figure}

The asymptotic $\mathbb{R}^2/\mathbb{Z}_2$ has the following interpretation: it describes the relative distance and angular orientation of the monopole pair in the center-of-mass frame. The reason for the $\mathbb{Z}_2$ is that the two monopoles are indistinguishable as classical field configurations. The scattering of monopoles in the $x$-$y$-plane is then described by a geodesic passing through the smoothed-out tip of the cone (see figure~\ref{fig:moduli-space} (left)). On the cone, it looks like we have changed the angle by $\pi$, but in terms of the coordinate $\varphi$, the change is only by $\pi/2$. Thus head-on colliding monopoles in the $x$-$y$-plane scatter at a 90-degree angle. The same explanation can be applied to the scattering of vortices, which also scatter by a 90-degree angle~\cite{Samols:1991ne}.

A second important geodesic surface is the `Atiyah-Hitchin trumpet' $\Sigma_2$ \cite{Atiyah-Hitchin1988}, see figure \ref{fig:moduli-space} (right). The asymptotic plane on the upper end corresponds to the $x$-$z$ scattering plane. For two monopoles scattering with zero impact parameter in that plane, the geodesic goes straight through the throat and reemerges on the other side. The plane on the other side describes two separate properties: the radial coordinate is the separation in the $x$-$z$-plane, while excursions in the angular direction generate an electric charge. Thus, in the scattering process, the monopoles do not stay in the same plane.

For sufficiently small impact parameters, this continues to be the case, even though the geodesic path will now wind around the cylindrical part. The monopoles will pick up an electric charge: we thus observe the scattering of magnetic monopoles into dyons~\cite{Atiyah:1985dv, Gibbons:1986df}. There is a critical value for the latter such that the geodesic line no longer traverses through the trumpet; in that case, it spends a long time in the throat, constituting a 2-monopole bound state, before decaying back into two monopoles. The scattering angle now depends on the value of the impact parameter. 

\section{\MakeUppercase{Limits of the Moduli Approximation}}
\label{sec:limits-of-the-moduli-approximation}
In this section, we briefly discuss the limitations of the moduli space approximation. 

At low velocities, $u$, the kinetic energy of a monopole (or a monopole pair) is $\sim M_\text{monopole} u^2$. The moduli space approximation predicts that after the scattering, the scattered monopoles have the same velocity. In particular, the moduli space approximation predicts zero energy loss into radiation. Since monopoles consist microscopically of charged $W$-bosons, this is obviously not exact: during the scattering process, there will be radiation emission $\sim v^5$\cite{Manton:1988bn}. 

One other source of deviation from the moduli space approximation is (classical) decoherence of the monopole profile when two or more of them are close to each other, of distance $\sim 1/m_v$. It has been estimated \cite{Stuart:1994tc} that such decoherence sets in when the monopoles are close to each other for times $t \gtrsim 1/v$. 

To summarize, the validity of the moduli space approximation demands that during the scattering process the inner structure 
of the monopoles is not perturbed significantly.

In principle, we can probe both of the above limitations in our simulation: we can choose relativistic initial velocities (see section \ref{sec:results}), and we can set up our initial monopoles in such a way that the moduli space approximation would predict the geodesic motion to spend a long time in the throat of the Atiyah-Hitchin trumpet. In such a case, we cannot expect the moduli space approximation to be precise, and we can probe deviations from it.
 This analysis is left for future work.

\section{\MakeUppercase{Charge-N Magnetic Monopoles}}
\label{sec:charge-n-magnetic-monopoles}
Once the magnetic monopole solution was discovered, a natural question arose: do magnetic monopoles with higher magnetic charges exist? Weinberg and Guth~\cite{Weinberg:1976eq} demonstrated that spherically symmetric monopoles with higher magnetic charges cannot have finite energies. However, non-spherically symmetric monopoles with multiple magnetic charges do exist at finite energy.
 While these configurations are unstable in the non-BPS case, they become stable solutions in the BPS limit.

\begin{figure*}
  \centering
  \begin{subfigure}{0.3\textwidth}
    \includegraphics[trim=0 0 0 0,clip,width=\textwidth]{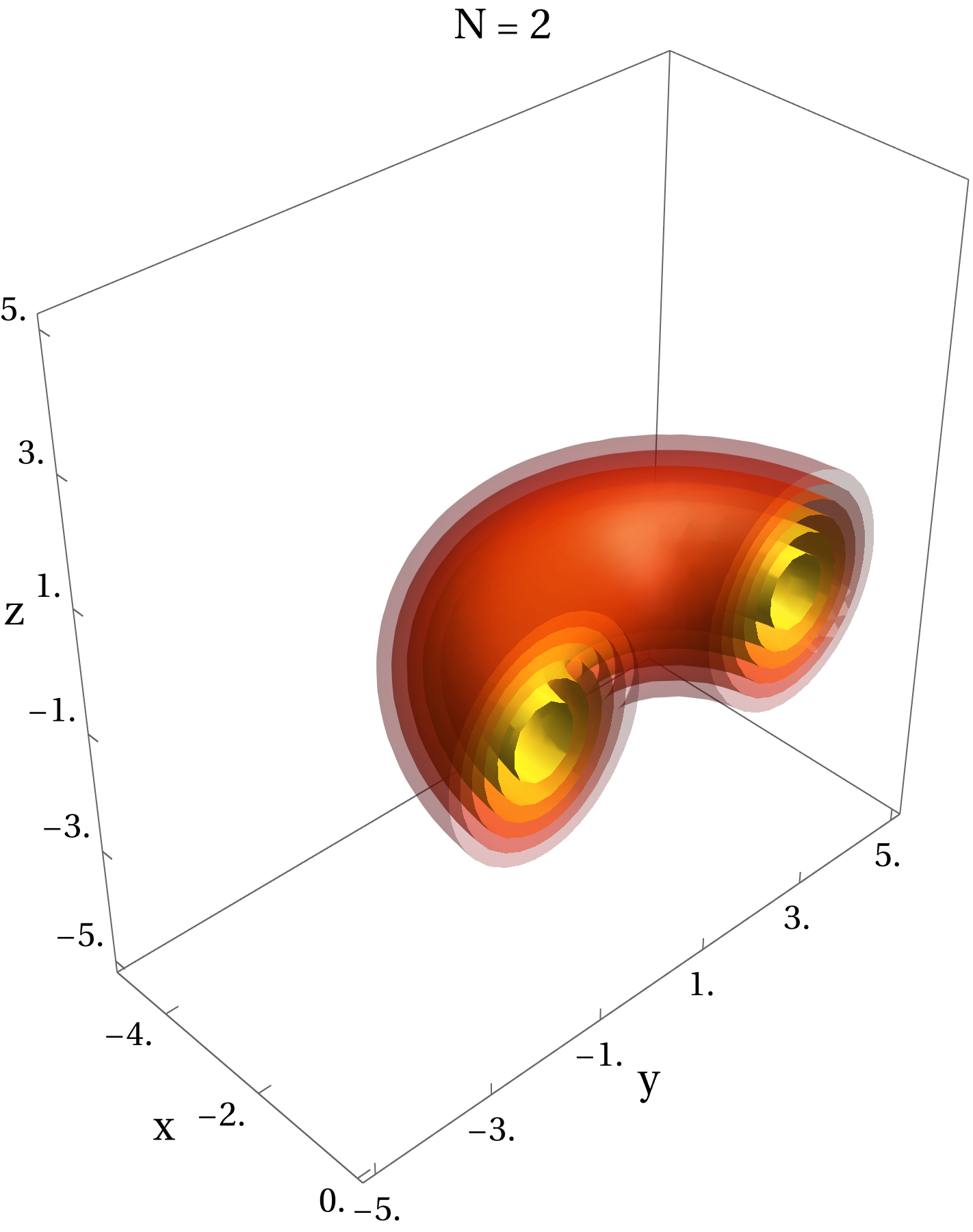}
  \end{subfigure}
  \hspace{\fill}
  \begin{subfigure}{0.3\textwidth}
    \includegraphics[trim=0 0 0 0,clip,width=\textwidth]{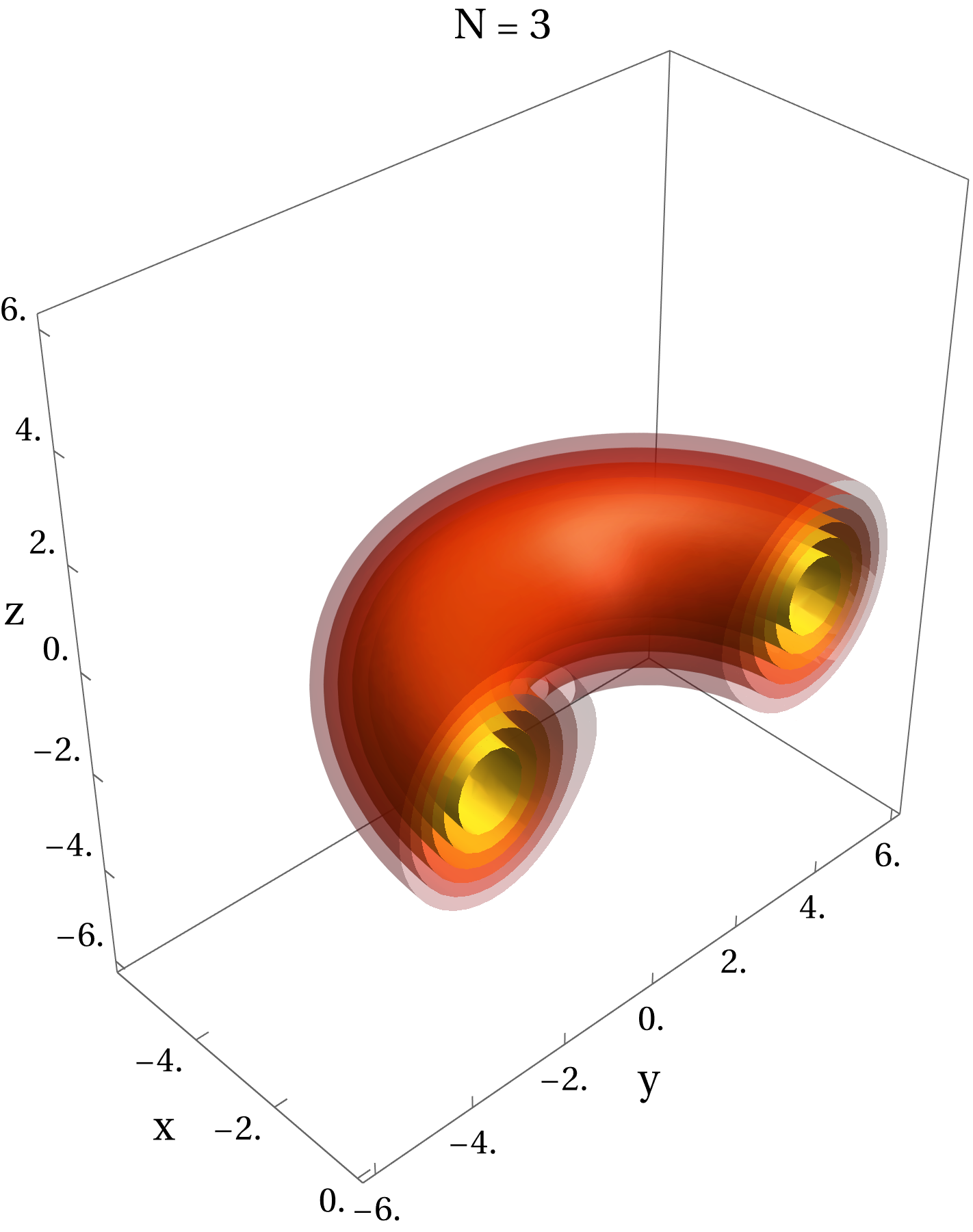}
  \end{subfigure}
  \hspace{\fill}
  \begin{subfigure}{0.3\textwidth}
    \includegraphics[trim=0 0 0 0,clip,width=\textwidth]{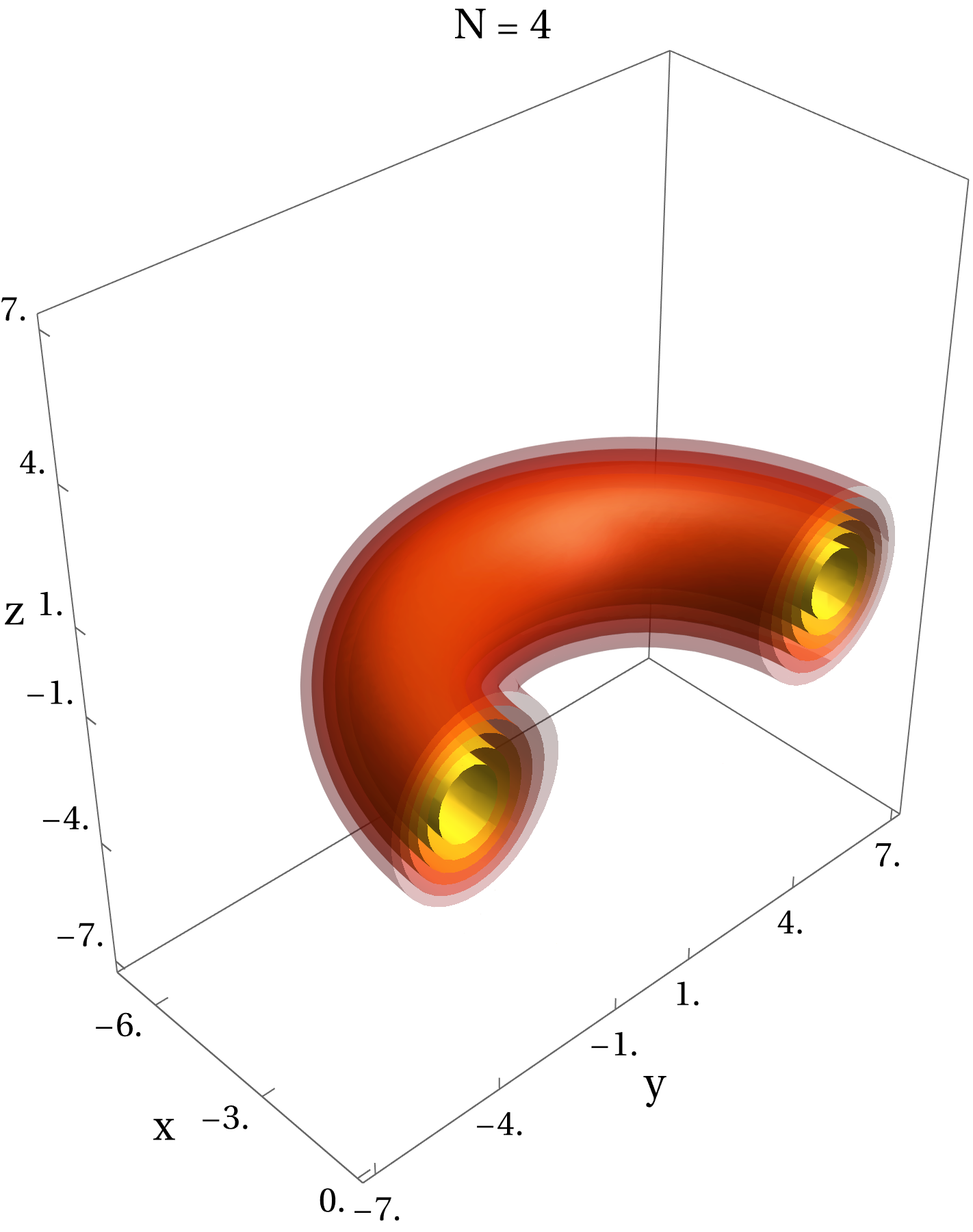}
  \end{subfigure}
  \caption{These contour plots illustrate the toroidal shape of the energy density for charge -$2$, -$3$, and -$4$ monopoles in the BPS limit $m_h \rightarrow 0$. We can observe that the radius of the torus increases for higher charges. The contours illustrate energy densities within the intervals $[0.045,0.07]$, $[0.032,0.053]$, and $[0.03,0.045]$ (in units of $m_v^4/g^2$), respectively. The length values are given in units of $m_v^{-1}$.}
  \label{fig:N-monopole-solution}
\end{figure*}

The simplest multiple-charge monopoles have toroidal symmetry~\cite{Ward:1981jb, FORGACS1981141, Prasad:1980hi}. We have obtained these solutions numerically using a relaxation technique described in the appendix. Concretely, we initialized the numerical iteration process with a configuration of the form
\begin{align}
    \hat{\phi}^1&=\cos{(N \varphi)}\sin{\theta},\nonumber\\
    \hat{\phi}^2&=\sin{(N \varphi)}\sin{\theta},\nonumber\\
    \hat{\phi}^3&=\cos{\theta}.
\end{align}
In order to satisfy $(D_i\phi)^a\rightarrow 0$ for $r\rightarrow \infty$, we chose the gauge field to be of the form
\begin{align}
    W^a_i\xrightarrow{r\rightarrow \infty} -\frac{1}{g}\varepsilon_{abc} \hat{\phi}^b \partial_i \hat{\phi}^c.
\end{align}
Figure~\ref{fig:N-monopole-solution} presents contour plots of the energy density for toroidal BPS monopoles with charges $2$, $3$, and $4$. It can be noticed that the radius of the torus increases as the charge increases.

\begin{figure}[h]
    \includegraphics[trim=0 0 0 0,clip,width=0.66\linewidth]{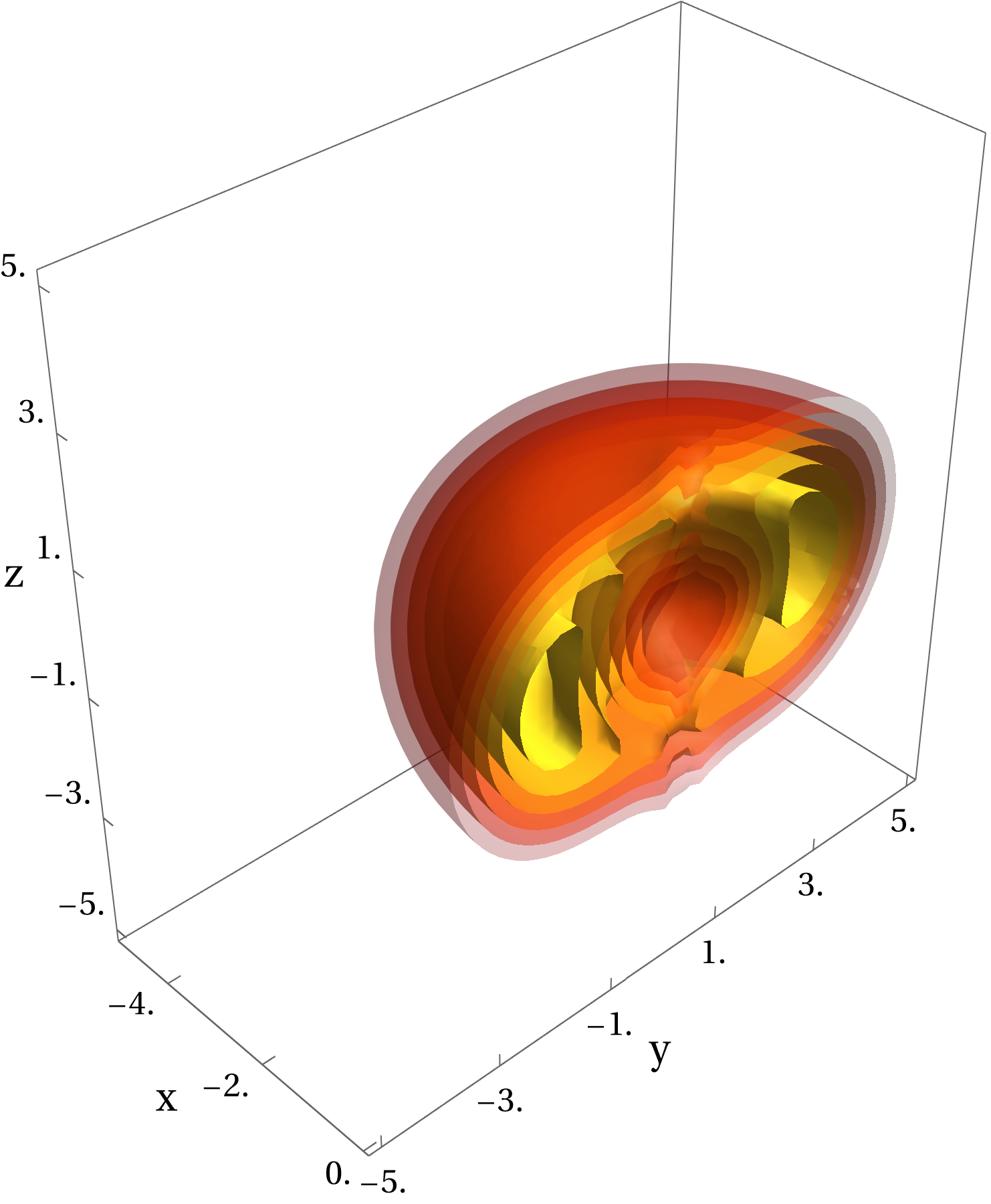}
    \caption{This contour plot illustrates the energy density within the interval $[0.04\, m_v^4/g^2, 0.065\, m_v^4/g^2]$ for a charge-$2$ magnetic monopole in the non-BPS case with $m_h/m_v=0.1$. The high-energy contours (yellow) form a toroidal shape. However, a low energy contribution (red) appears in the center of the torus, which we didn't observe in the BPS case, shown in Figure~\ref{fig:N-monopole-solution}. The coordinates are given in units of $m_v^{-1}$.}
    \label{fig:2Monopole-nonBPS}
\end{figure}

Using the numerical relaxation method, we can also identify unstable solutions for higher-charge monopoles in the non-BPS case. The energy density of the charge-$2$ monopole for $m_h/m_v=0.1$ is shown in Figure~\ref{fig:2Monopole-nonBPS}. We observe that the toroidal shape is spoiled by a small energy contribution at the center of the torus. For higher values of $m_h/m_v$, the toroidal structure vanishes entirely, and the object becomes smaller because the energy is concentrated within a radius of $\sim m_h^{-1}$. Notice that these solutions for $m_h \neq 0$ are unstable and would rapidly decay into two separate monopoles with unit winding numbers.

\section{\MakeUppercase{Two-Monopole Configuration}}
\label{sec:two-monopole-configuration}
An exact analytical solution for two separated BPS magnetic monopoles with the same charge is not known. However, we have identified two approximate analytical solutions that describe this configuration rather well.

\textit{Monopoles located on the $x$-axis.}
The first ansatz describes two monopoles located in the $x$-$y$-plane. This ansatz was inspired by the charge $2$ magnetic monopole configuration discussed in the previous section. Instead of using $2\varphi$, we replaced it with $\varphi_1 + \varphi_2$, where $\varphi_1$ and $\varphi_2$ represent the azimuthal angles around the positions of the first and second monopole, respectively. The resulting ansatz is

\begin{align}
    \hat{\phi}^1&=\cos(\varphi_1+\varphi_2)\sin\theta,\nonumber\\
    \hat{\phi}^2&=\sin(\varphi_1+\varphi_2)\sin\theta,\nonumber\\
    \hat{\phi}^3&=\cos{\theta},
\end{align}
with
\begin{align*}
    \varphi_{1,2}&=\text{arctan2}(y,x-x_{1,2})\\
    \theta&=\arccos{\left(\frac{z}{r_1}\right)}\Theta_H(-x)+\arccos{\left(\frac{z}{r_2}\right)}\Theta_H(x)\\
    r_{1,2}&=\sqrt{(x-x_{1,2})^2+y^2+z^2},
\end{align*}
where $\Theta_H$ is the Heaviside step function, and $x_1=-x_2<0$.

\begin{figure}[h]
    \includegraphics[trim=0 0 0 0,clip,width=\linewidth]{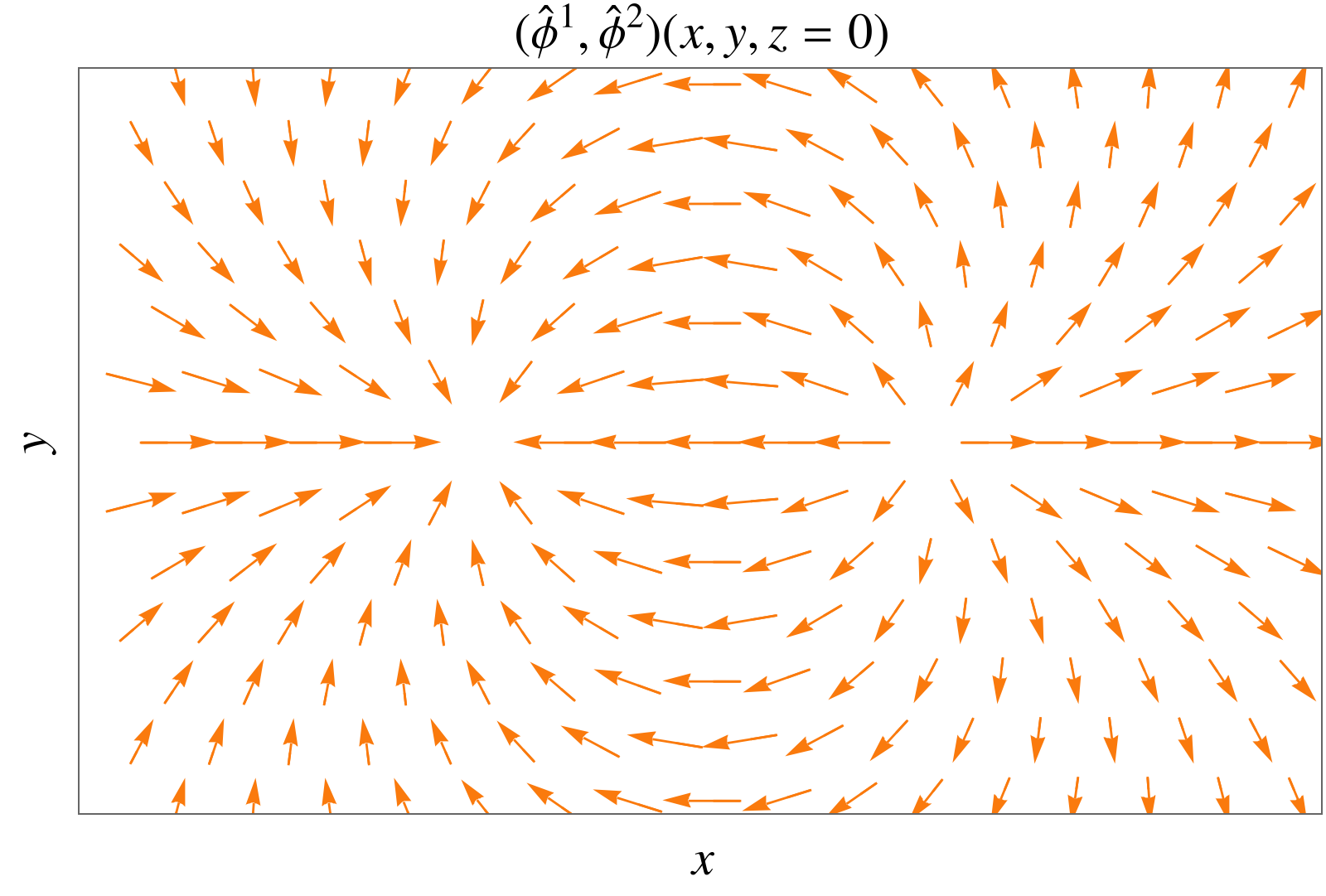}
    \includegraphics[trim=0 0 0 0,clip,width=\linewidth]{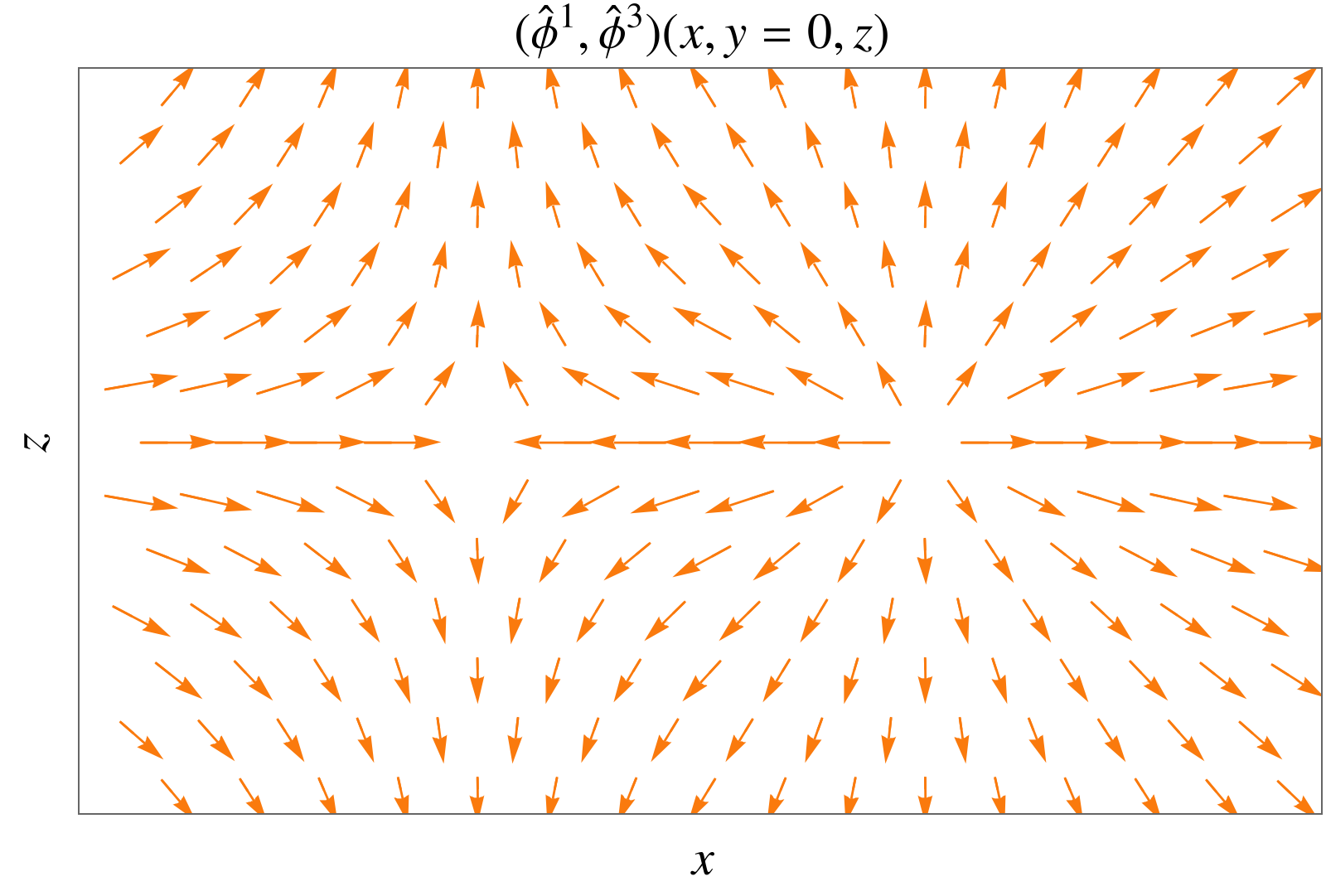}
    \caption{The scalar field direction $(\hat{\phi}^1, \hat{\phi}^2)^T$ in the $x$-$y$-plane (top) and $(\hat{\phi}^1, \hat{\phi}^3)^T$ in the $x$-$z$-plane (bottom) is illustrated in a vector plot.}
    \label{fig:scalar-field-direction}
\end{figure}

Figure~\ref{fig:scalar-field-direction} illustrates the vector plot for $\hat{\phi}^a$ in the $x$-$y$- and $x$-$z$-planes. From the figure, it is clear that this configuration has no axial symmetry around the $x$-axis. This is in contrast with a monopole-antimonopole configuration, which exhibits axial symmetry.

Since the direction of the scalar field is determined, we can write down the complete ansatz in terms of the profile functions
\begin{align}
    \phi^a&=\frac{1}{m_v g}\frac{H(r_1)}{r_1}\frac{H(r_2)}{r_2}\, \hat{\phi}^a,\\
    W^a_i&=-\frac{1}{g}(1-K(r_1))(1-K(r_2))\, \varepsilon_{abc} \hat{\phi}^b \partial_i \hat{\phi}^c,
\end{align}
where $H(r)$ and $K(r)$ are given by the BPS solutions~\eqref{eq:BPS-solution}.

\textit{Monopoles located on the $z$-axis.}
Sometimes it could be useful to have the two monopoles located on the $z$-axis. For instance, in scenarios where monopoles are connected by strings, the solutions are usually configured so that the string is located along the $z$-axis.

For this solution, we used a monopole-antimonopole configuration as a basis. As shown in~\cite{Saurabh:2017ryg}, such a configuration includes a twist as an additional degree of freedom. The analytic monopole-antimonopole ansatz with maximal twist produces a magnetic field with a repulsive shape between the two monopoles. However, a sign flip of the magnetic field occurs at the central plane between the two monopoles. We made use of this property and compensated the sign flip, by changing the sign of $\varphi$ at the central plane.

The final ansatz is
\begin{align}
    \hat{\phi}^1&=\cos{\varphi}\sin{(\theta_1 + \theta_2)},\nonumber\\
    \hat{\phi}^2&=\sin{\varphi}\sin{(\theta_1 + \theta_2)},\nonumber\\
    \hat{\phi}^3&=\cos{(\theta_1 + \theta_2)},
\end{align}
with
\begin{align*}
    \varphi&=\text{sign}(z)\, \text{arctan2}(y,x)\\
    \theta_{1,2}&=\arccos{\left(\frac{z-z_{1,2}}{r_{1,2}}\right)}\\
    r_{1,2}&=\sqrt{x^2+y^2+(z-z_{1,2})^2},
\end{align*}
where $z_1=-z_2$.

Numerical simulations revealed that this ansatz is less precise than the previous one. Therefore, in the following discussion, we will focus on the configuration with two monopoles located on the $x$-axis.

\section{\MakeUppercase{N-Monopole Configuration}}
\label{sec:n-monopole-configuration}
The two-monopole configuration can be generalized to $N$ monopoles arranged in a cyclically symmetric way within a single plane. We start again with a $N$ monopole configuration and split $N \phi$ into a sum of $N$ distinct angles $\phi_n$, each describing the azimuthal angle around the position of one of the $N$ monopoles.
Consequently, the normalized scalar field vector is:
\begin{align}
    \hat{\phi}^1&=\cos\left(\sum_n \varphi_n\right)\sin\theta,\nonumber\\
    \hat{\phi}^2&=\sin\left(\sum_n \varphi_n\right)\sin\theta,\nonumber\\
    \hat{\phi}^3&=\cos{\theta},
\end{align}
with
\begin{align}
    \varphi_n=&\text{arctan2}(y_n,x_n),\nonumber \\
    \theta=&\sum_n \arccos{\left(\frac{z}{r_n}\right)}\nonumber\\
    &\hspace{0.5cm}\cdot \Theta_H \left( \cos{\left(\alpha_n - \Delta \alpha/2\right)}\, x+\sin{\left(\alpha_n - \Delta \alpha /2\right)}\, y \right)\nonumber\\
    &\hspace{0.5cm}\cdot \Theta_H \left( \cos{\left(\alpha_n + \Delta \alpha/2\right)}\, x+\sin{\left(\alpha_n + \Delta \alpha /2\right)}\, y \right),\nonumber\\
    r_n=&\sqrt{x_n^2+y_n^2+z^2}.
\end{align}
The coordinates $x_n$ and $y_n$ are the $x$ and $y$ positions of the monopoles. In order to arrange the monopoles in cyclic symmetry, we write these coordinates as
\begin{align} 
\label{eq:coordinates-n-monopole}
    x_n &= \cos(\alpha_n)x + \sin(\alpha_n)y - R, \nonumber\\
    y_n &= \cos(\alpha_n)y - \sin(\alpha_n)x, 
\end{align}
where $R$ is the distance from the monopoles to the center of the configuration, and $\alpha_n$ is the angle between the $x$-axis and the position of the $n$th monopole. The angle between two neighboring monopoles is $\Delta \alpha$.
For a configuration with three monopoles, we set $\alpha_n = \lbrace \frac{\pi}{3}, \pi, \frac{5\pi}{3} \rbrace$, giving $\Delta \alpha = \frac{2\pi}{3}$. For four monopoles, we choose $\alpha_n = \lbrace 0, \frac{\pi}{2}, \pi, \frac{3\pi}{2} \rbrace$, resulting in $\Delta \alpha = \frac{\pi}{2}$.\\

\section{\MakeUppercase{Results of the Simulations and Discussion}}
\label{sec:results}
\begin{figure*}
  \centering
  \begin{subfigure}{0.3\textwidth}
    \includegraphics[trim=0 0 0 0,clip,width=\textwidth]{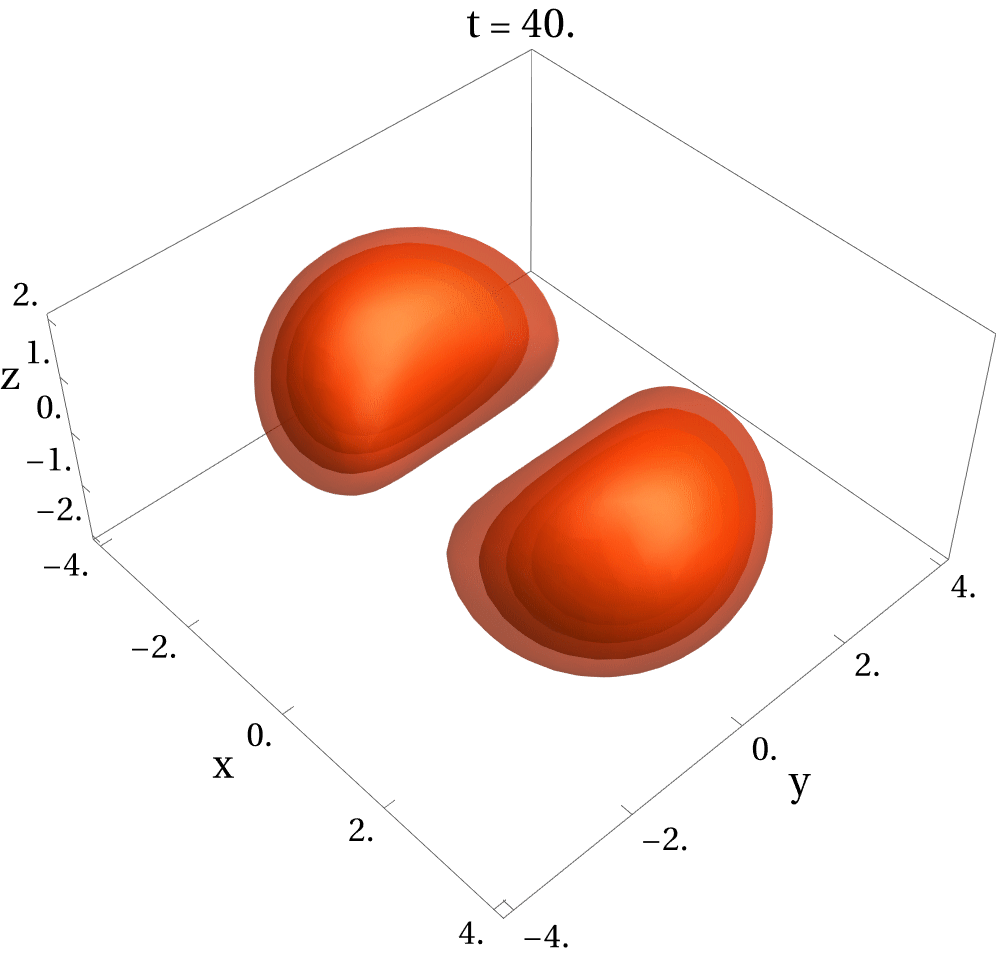}
  \end{subfigure}
  \hspace{\fill}
  \begin{subfigure}{0.3\textwidth}
    \includegraphics[trim=0 0 0 0,clip,width=\textwidth]{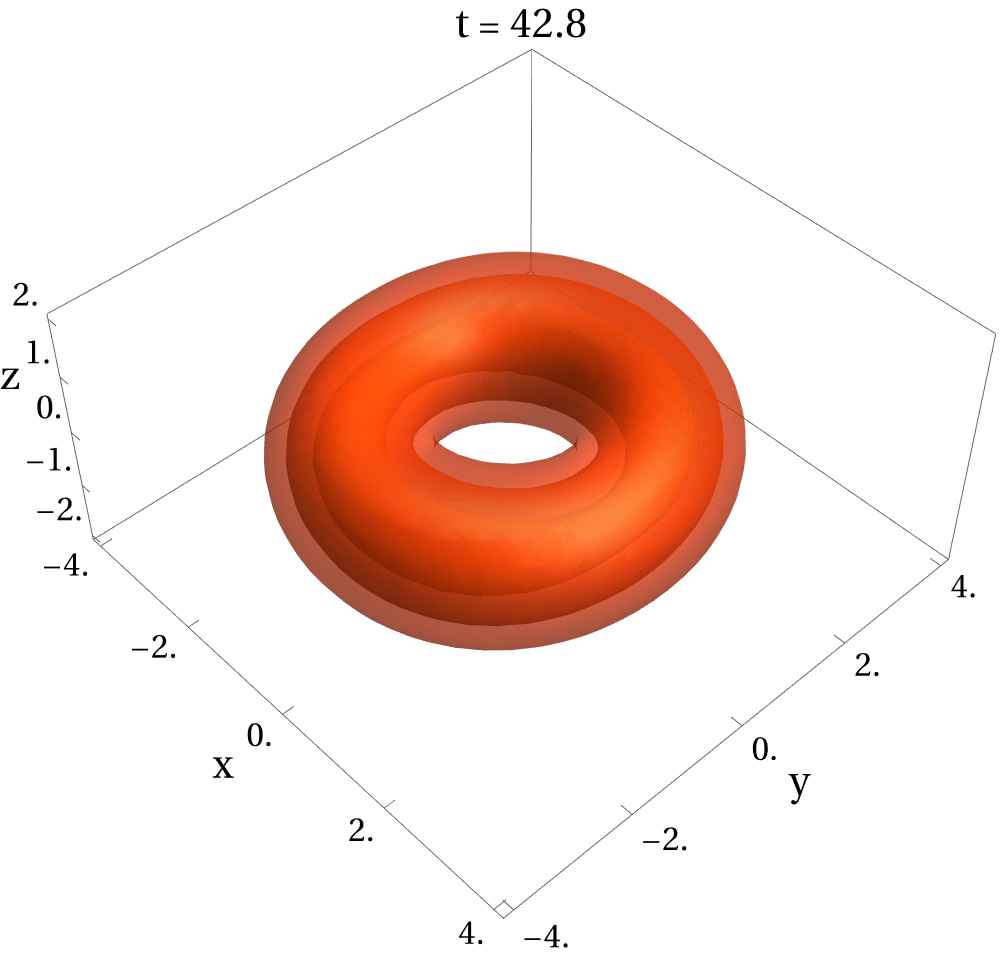}
  \end{subfigure}
  \hspace{\fill}
  \begin{subfigure}{0.3\textwidth}
    \includegraphics[trim=0 0 0 0,clip,width=\textwidth]{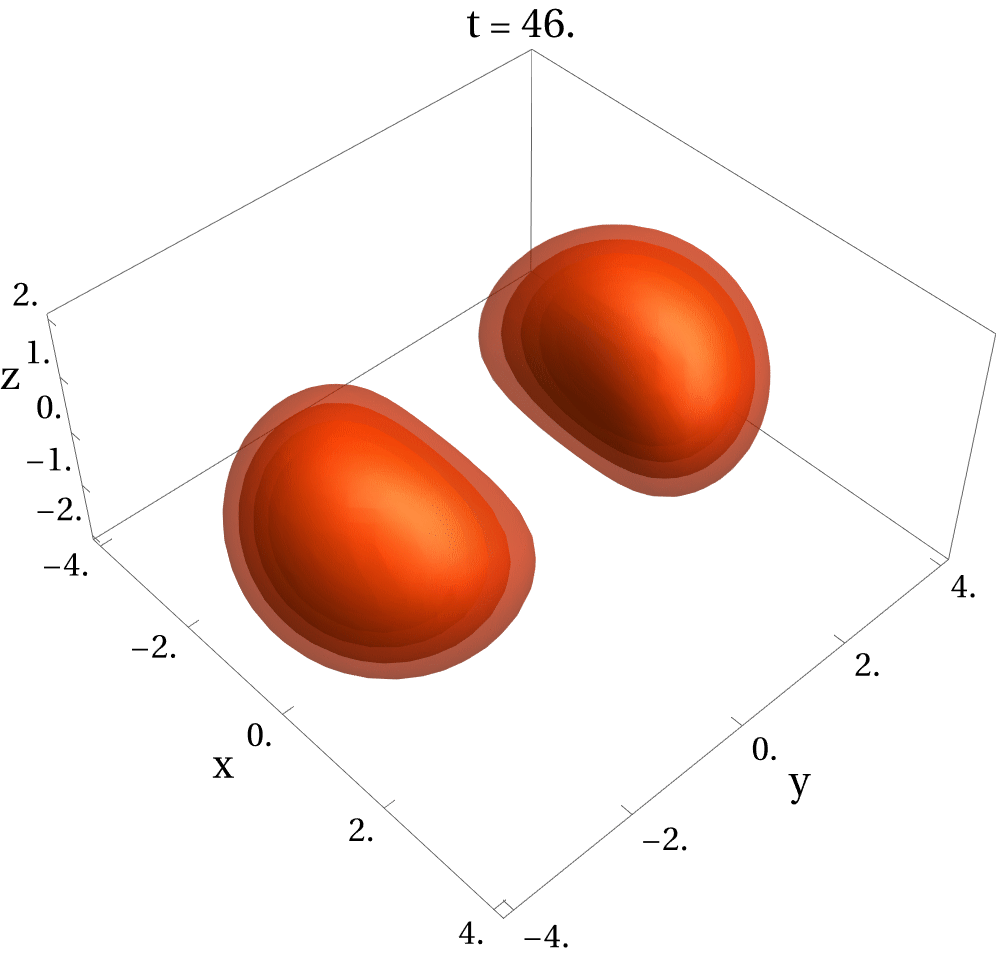}
  \end{subfigure}
  \caption{These contour plots show three frames for the energy densities bigger than $0.06 m_v^4/g^2$ for the two monopoles right-angle scattering. As we can see, the energy density forms a toroidal structure during the scattering process.  The length and time values are given in units of $m_v^{-1}$.}
  \label{fig:2monopole-scattering}
\end{figure*}

We simulated the head-on two-monopole scattering for several cases. Initially, the monopole velocities were set to $0.2$ (in units of the speed of light). The simulation showed the right-angle scattering in the $x$-$y$-plane, with the two monopoles forming a toroidal state during the collision. These results are in complete agreement with the predictions of the moduli approximation~\cite{Atiyah-Hitchin1988}. Figure~\ref{fig:2monopole-scattering} shows some frames of the scattering process. The results of the simulations can also be found in the following video:\\
\url{https://youtu.be/a2wOOo2AAHI} .

When the monopoles are far apart, the energy density appears to be distributed cylindrically symmetric. This might suggest that the configuration should preserve this symmetry, preventing the monopoles from scattering in any specific direction. However, as the monopoles approach each other, the energy distribution forms a toroidal shape. This change confirms that the configuration does not maintain cylindrical symmetry at large distances.
Although this might not be visible from the energy density alone, it can be observed in the internal structure of the scalar field: as discussed in section~\ref{sec:two-monopole-configuration}, the scalar field $\phi^a$ is not axially symmetric.

From this vector plot, an analogy to the scattering of vortices can be found. As discussed in~\cite{Shellard:1988zx,Samols:1991ne}, vortices also scatter with $90$ degrees. Plotting the complex scalar field of the vortex $\psi = \Re{\psi} + i \Im{\psi}$ as a vector plot produces a pattern similar to that shown in Figure~\ref{fig:scalar-field-direction} (top). Therefore, the scattering of monopoles in the $x$-$y$-plane shares a resemblance with the scattering of two vortices.
Similarly, the scalar field direction in the $x$-$z$-plane, shown in Figure~\ref{fig:scalar-field-direction} (bottom), exhibits a field direction akin to a vortex-antivortex configuration. In vortex-antivortex scattering, the vortex and the antivortex annihilate, and their winding numbers cancel. Although the monopoles do not annihilate, in the $x$-$z$-plane, the $2D$ winding number vanishes similarly to the scattering of a vortex with an antivortex. For the monopoles, this means that they leave the $x$-$z$-plane, i.e. they move apart along the $y$-direction.\\

\begin{figure}[h]
    \includegraphics[trim=0 0 0 15,clip,width=\linewidth]{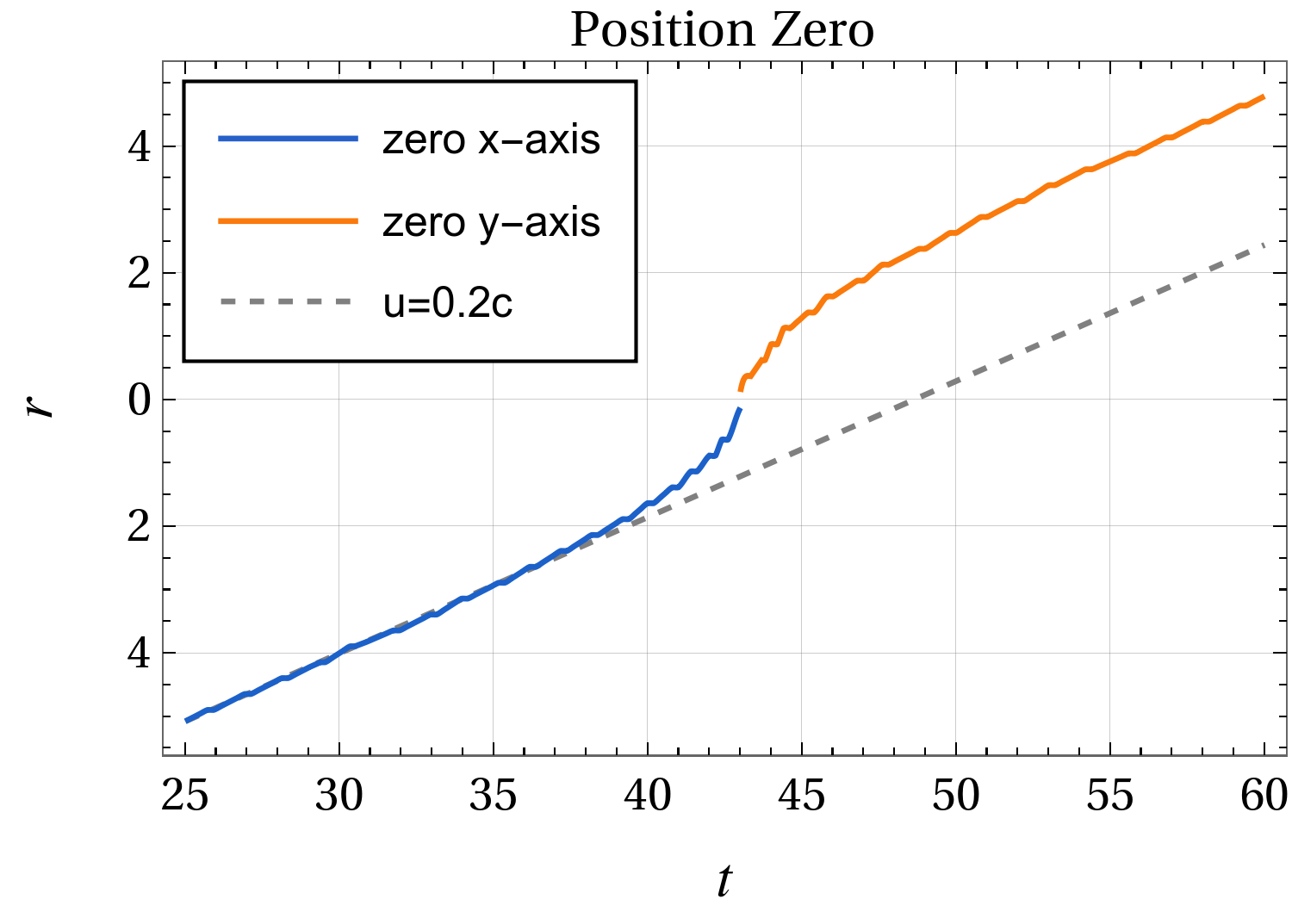}
    \caption{This plot shows the distance between the zero of the scalar field $\phi^a=0$ and the origin. Initially, the zeros move along the $x$-axis (blue) and after the $90$ degree scattering, the zeros move along the $y$-axis. The length and time are given in units of $m_v^{-1}$.}
    \label{fig:position-zeros}
\end{figure}

Using the data from our simulations, we tracked the zeros of the scalar field $\phi^a = 0$. At large separations, these zeros coincide with the centers of the magnetic monopoles. Thus, tracking the zeros allows us to determine the velocities of the monopoles before and after the collision. In Figure~\ref{fig:position-zeros}, we show the trajectory of the zeros for initial monopole velocities $u = 0.2$. We observe that the monopole velocities before and after the collision remain nearly unchanged, with a velocity change of about $\Delta u = 0.01$. In a relativistic case with an initial velocity of $u = 0.62$, the change is slightly larger with $\Delta u = 0.04$, corresponding to roughly $5$ percent of the monopole mass being emitted as radiation. Thus, we conclude that radiation emission during the scattering process is minimal. This observation aligns with the predictions of Manton and Samols, who showed in~\cite{Manton:1988bn} that radiation emission is rather small because the configuration has no dipole moment and the dominant contribution arises from the quadrupole order.

When the monopoles approach each other, the energy density distributes into a toroidal shape. During this torus formation, the zeros associated with the monopoles leave their cores and converge at the origin. This is the only possibility through which an intermediate two-monopole state can form~\cite{Ward:1981jb}. At the origin, the zeros scatter at a right angle before returning to the cores of the two monopoles. 
We observed that the velocities of the zeros can exceed the speed of light. This observation is fully consistent with special relativity, as the zeros do not carry any physical quantities such as energy.
Correspondingly, they cannot transmit any information.
Instead, all the energy is propagated along the torus and moves at speeds below the speed of light.

As discussed in section~\ref{sec:limits-of-the-moduli-approximation}, the moduli approximation is valid only in the BPS limit and for small velocities. Our numerical simulations, however, enabled us to explore monopole scattering beyond this approximation. In particular, we analyzed scenarios with relativistic velocities, examining cases with velocities up to $u=0.8$. In all these cases, we observed the characteristic right-angle scattering.

\begin{figure}[h]
    \includegraphics[trim=0 0 0 0,clip,width=0.66\linewidth]{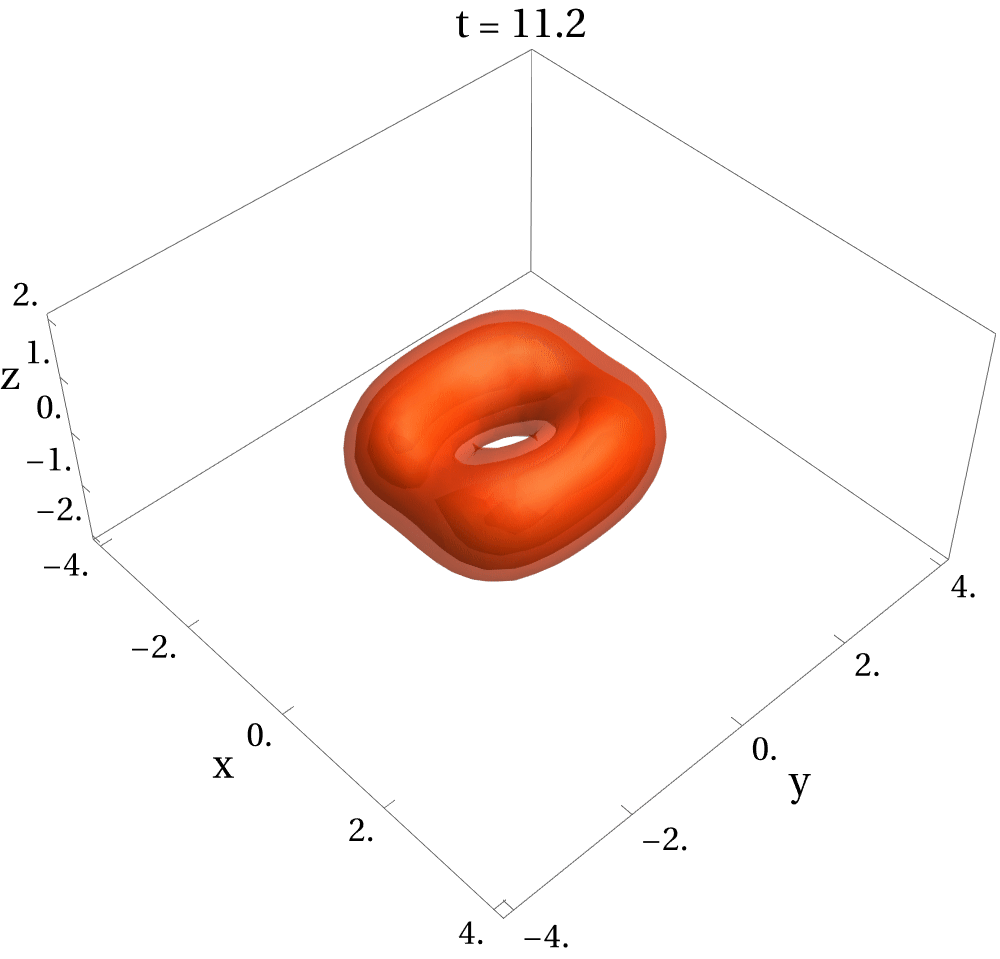}
    \caption{This contour plot shows the energy densities bigger than $0.015 m_v^4/g^2$ for the two monopoles right-angle scattering with relativistic velocity $u=0.6$. In this high-velocity regime, the shape was never an axially symmetric torus. The length and time values are given in units of $m_v^{-1}$.}
    \label{fig:monopole-scattering-relativistic}
\end{figure}

We noticed that the monopoles do not form a perfectly axially symmetric intermediate state during the collision. Instead, the toroidal structure becomes slightly deformed, with straightened edges, and the 'radius' is smaller, as shown in Figure~\ref{fig:monopole-scattering-relativistic}. This deformation arises because the individual monopoles are no longer spherically symmetric when they collide. Due to Lorentz contraction, the energy density of a relativistically moving monopole is distributed in an ellipsoidal shape.

Furthermore, we utilized our simulations for the analysis of  the scattering of monopoles in non-BPS cases ($m_h \neq 0$). It is important to note that in this regime, the scalar force is exponentially damped
at distances $r > m_h^{-1}$. As a result, the magnetic repulsion dominates, creating an effective barrier that must be overcome for the monopoles to approach each other sufficiently closely 
for the scalar attraction to become significant.

 In order to estimate the required energy, we treat the monopoles, initially placed far away from each other, as point particles.  
 The corresponding repulsive magnetic potential is given by
\begin{align*}
    \Delta E_\text{magn}\sim\frac{q_m^2}{r}\sim \frac{1}{g^2 r}.
\end{align*}
The scalar interactions become relevant at distances of 
order $r \sim m_h^{-1}$.    Correspondingly,  the energy required for overcoming  the magnetic potential barrier is 
\begin{align*}
    \Delta E_\text{magn}\sim \frac{m_h}{g^2} \sim \frac{m_h}{m_v} M_\text{monopole}.
\end{align*}
Therefore, the Lorenz factor that is necessary for  bringing the monopoles sufficiently close for a possible $90$ degree scattering is of order
\begin{align}
    \gamma-1 \sim \frac{m_h}{m_v}.
\end{align}
These estimates hold for $m_v \gtrsim m_h$. For $m_v \ll m_h$, two distinct regimes emerge. At distances $r \gg m_v^{-1}$, the monopole interaction remains Coulomb-like. However, once the monopoles enter the 'gauge umbrella' at $r \sim m_v^{-1}$, the repulsive force becomes constant, similar to the situation with global monopoles\footnote{The connection with global monopoles is clear 
since this regime can be thought of as the $g \rightarrow 0$ limit of the theory. In this case, the repulsive constant force is due to the winding of the Goldstone phase.}. 
Consequently, bringing the monopoles closer together requires an immense amount of energy ($\sim \frac{m_h^2}{m_v^2} M_\text{monopole}$).

 Of course, the above qualitative arguments neglect several details, such as the deformation of the monopoles into a different energy shape (described in section~\ref{sec:charge-n-magnetic-monopoles}) and the role of non-linear scalar interactions. Nonetheless, our qualitative numerical analysis indicates that for $m_h/m_v\in [0, 1.0]$, velocities below $u=0.8$ are sufficient for achieving the right-angle scattering. We also performed simulations for global monopoles and observed that the repulsion between them is exceedingly strong, preventing the observation of close-range monopole scattering.\\

We also studied the planar scattering of three and four monopoles arranged in a cyclic symmetric way. As expected, the three monopoles scatter with $60$ degrees, and the four monopoles scatter with $45$ degrees. Again the monopoles propagate through a toroidal configuration when they come close to each other. The radius of the torus is larger than in the two-monopole case, as previously noticed in section~\ref{sec:charge-n-magnetic-monopoles}. The animations corresponding to the three and four monopoles scattering can be found in the following video:\\
\url{https://youtu.be/a2wOOo2AAHI}.

Again, for higher velocities, a deformation of the toroidal structure was observed as already explained previously for the two monopole scenario.

\section{\MakeUppercase{Non-planar scattering}}
\label{sec:non-planar-scattering}
So far, we have described the scattering of monopoles within a single plane, starting with monopole motion in the $x$-$y$-plane and ending with monopoles remaining in the $x$-$y$-plane after scattering. However, this is not the only possible outcome of an $N$-monopole scattering process. Depending on the orientation of the fields, non-planar scattering can occur, with monopoles acquiring momentum in the $z$-direction.

Using rational maps, Hitchin, Manton, and Murray showed in~\cite{Hitchin:1995qw} that three monopoles arranged 
cyclic-symmetrically can scatter into a single monopole moving in the positive $z$-direction and a toroidal monopole of charge-two moving in the negative $z$-direction, or vice versa. During the interaction, when the monopoles come close to each other, they pass through a tetrahedral monopole configuration.

A non-planar scattering was also analyzed for four monopoles in~\cite{Hitchin:1995qw}. In this case, similar dynamics occur. Four monopoles can scatter into a single monopole and a toroidal monopole of charge-three. Alternatively, the four monopoles can scatter into two toroidal monopoles with charges two. During this process, the configuration passes through a cubic monopole state. The tetrahedral and cubic monopole configurations are discussed in more detail in~\cite{Houghton:1995bs}.

We investigated these types of scatterings by starting with their outcomes. Specifically, we considered a toroidal monopole of charge $N_1$ colliding with a toroidal monopole of charge $N_2$ along the $z$-axis. After the collision, they separate into $N_1 + N_2$ charge-one monopoles.

For the initial configuration in the simulation, the monopoles are positioned on the $z$-axis at $z_1$ and $z_2$, described by $\phi_1^a$ and $\phi_2^a$, respectively. These configurations are then combined at the $z=0$ plane using
\begin{align}
    \hat{\phi}^a=\frac{1}{1+e^{\beta m_v z}}\, \hat{\phi_1}^a+\frac{1}{1+e^{-\beta m_v z}}\, \hat{\phi_2}^a.
\end{align}
In our simulation, we chose $z_1=-20 m_v^{-1}$, $z_1=20 m_v^{-1}$, and $\beta=0.5$.

The scalar field directions $\hat{\phi_1}^a$ and $\hat{\phi_2}^a$ are given by
\begin{align}
    \hat{\phi_1}^1 &=\cos(N_1 \varphi)\sin\theta_1, &&\hat{\phi_2}^1=-\cos(N_2 \varphi)\sin\theta_2,\nonumber\\
    \hat{\phi_1}^2 &=\sin(N_1 \varphi)\sin\theta_1, && \hat{\phi_2}^2 =\sin(N_2 \varphi)\sin\theta_2,\nonumber\\
    \hat{\phi_1}^3 &=\cos \theta_1, &&\hat{\phi_2}^3 =-\cos \theta_2,
\end{align}
with
\begin{align}
    \varphi&=\text{arctan2}(y,x),\nonumber\\
    \theta_{1,2}&=\arccos \left(\frac{z-z_{1,2}}{r_{1,2}}\right),\nonumber\\
    r&=\sqrt{x^2+y^2+(z-z_{1,2})^2}.
\end{align}
Note that the minus signs in $\hat{\phi}_2^a$ are essential for ensuring the correct orientation at the central $z=0$ plane between the two monopoles.

\begin{figure*}
  \centering
  \begin{subfigure}{0.3\textwidth}
    \includegraphics[trim=0 0 0 0,clip,width=\textwidth]{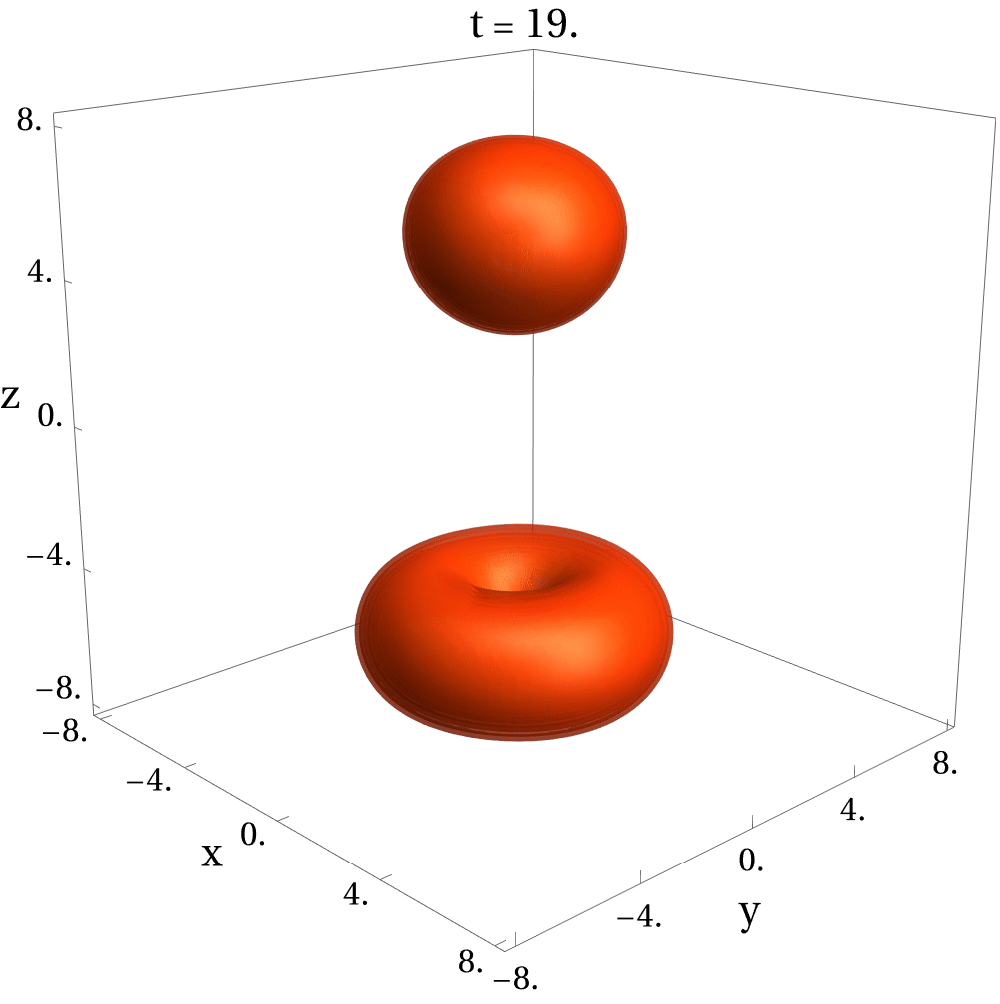}
  \end{subfigure}
  \hspace{\fill}
  \begin{subfigure}{0.3\textwidth}
    \includegraphics[trim=0 0 0 0,clip,width=\textwidth]{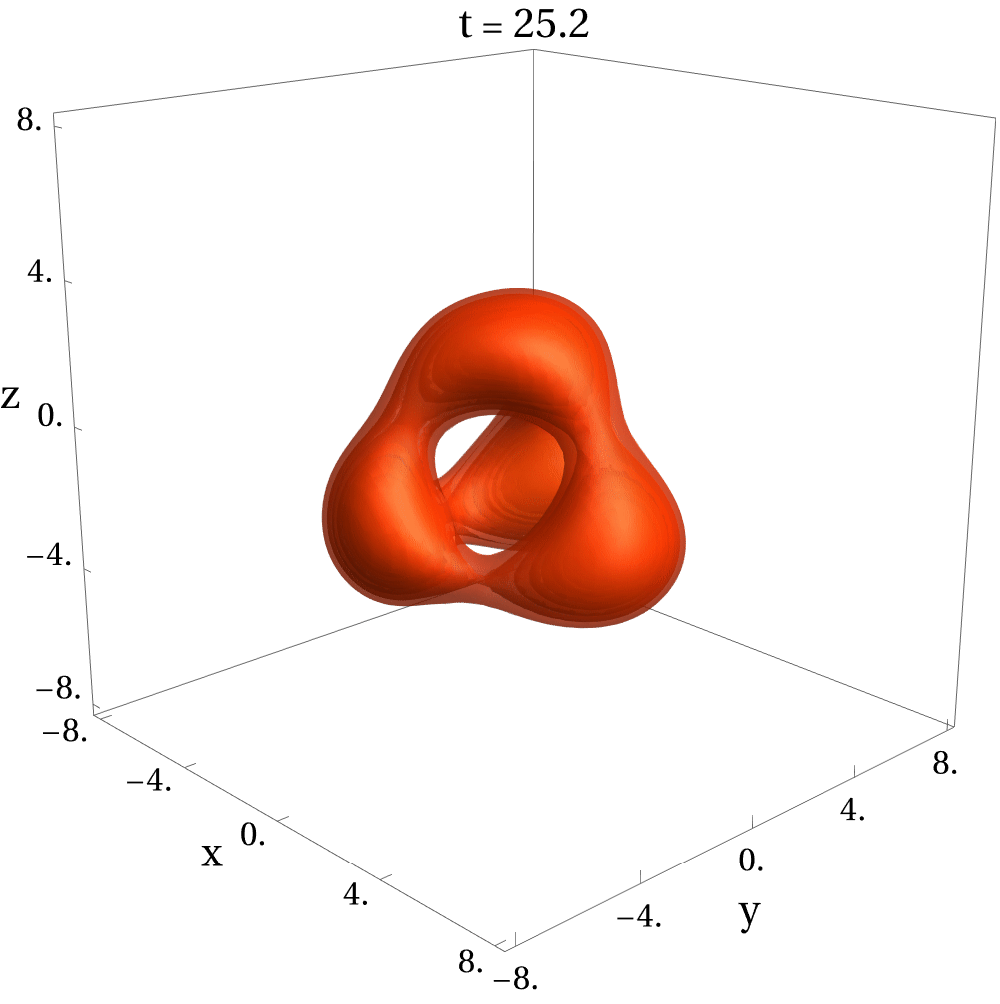}
  \end{subfigure}
  \hspace{\fill}
  \begin{subfigure}{0.3\textwidth}
    \includegraphics[trim=0 0 0 0,clip,width=\textwidth]{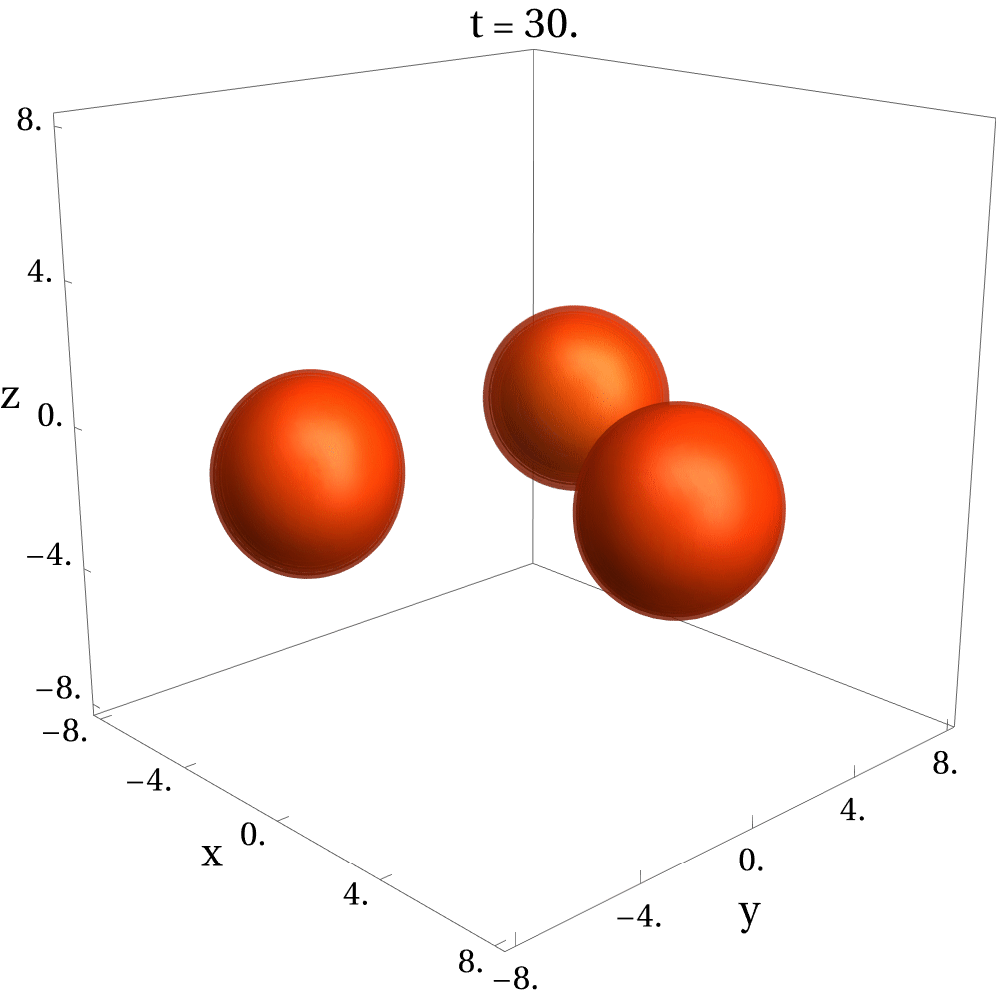}
  \end{subfigure}
  \caption{These contour plots show three frames for the energy densities bigger than $0.025 m_v^4/g^2$ for a toroidal charge-$2$ monopole colliding with a single monopole. Before the monopoles split into three monopoles the energy density forms a tetrahedron. The length and time values are given in units of $m_v^{-1}$.}
  \label{fig:21monopole-scattering}
\end{figure*}

\begin{figure*}
  \centering
  \begin{subfigure}{0.3\textwidth}
    \includegraphics[trim=0 0 0 0,clip,width=\textwidth]{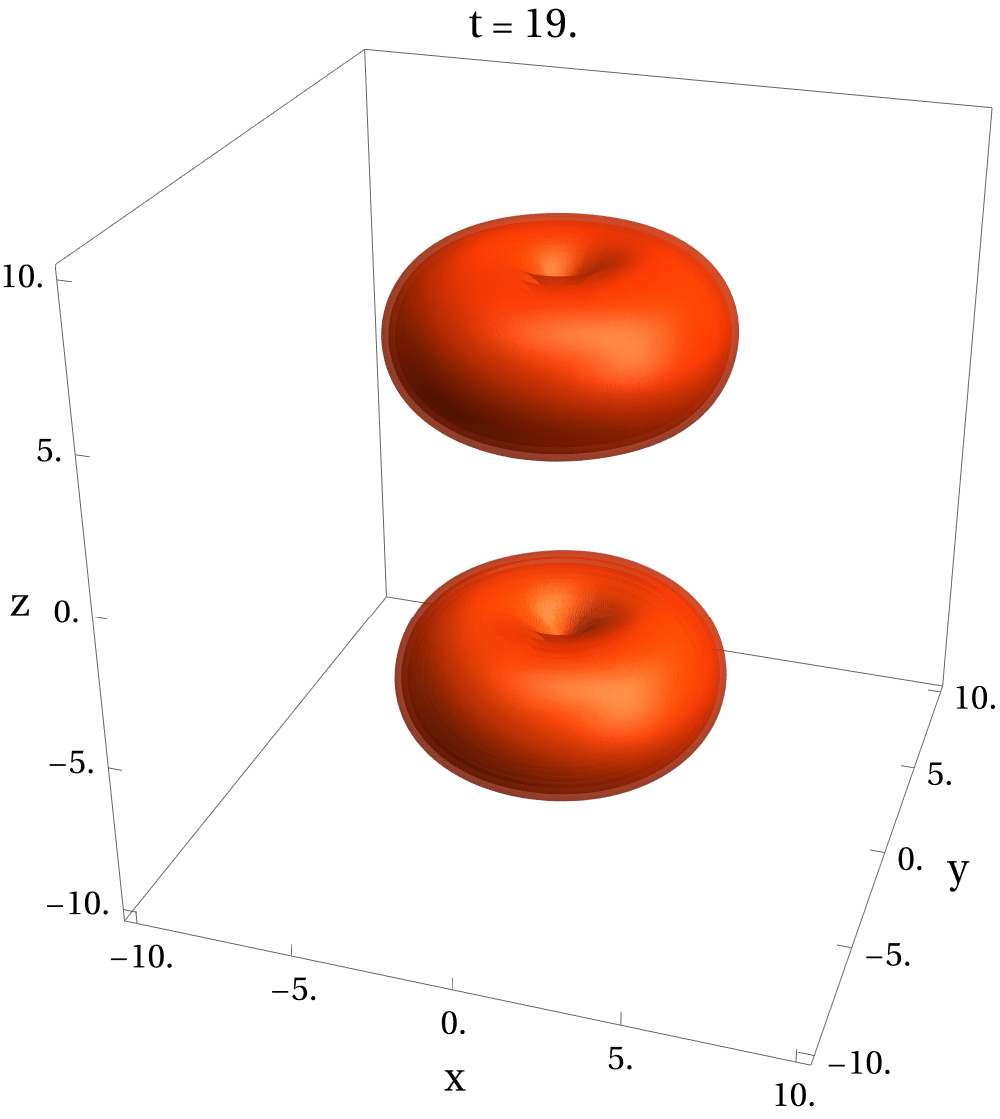}
  \end{subfigure}
  \hspace{\fill}
  \begin{subfigure}{0.3\textwidth}
    \includegraphics[trim=0 0 0 0,clip,width=\textwidth]{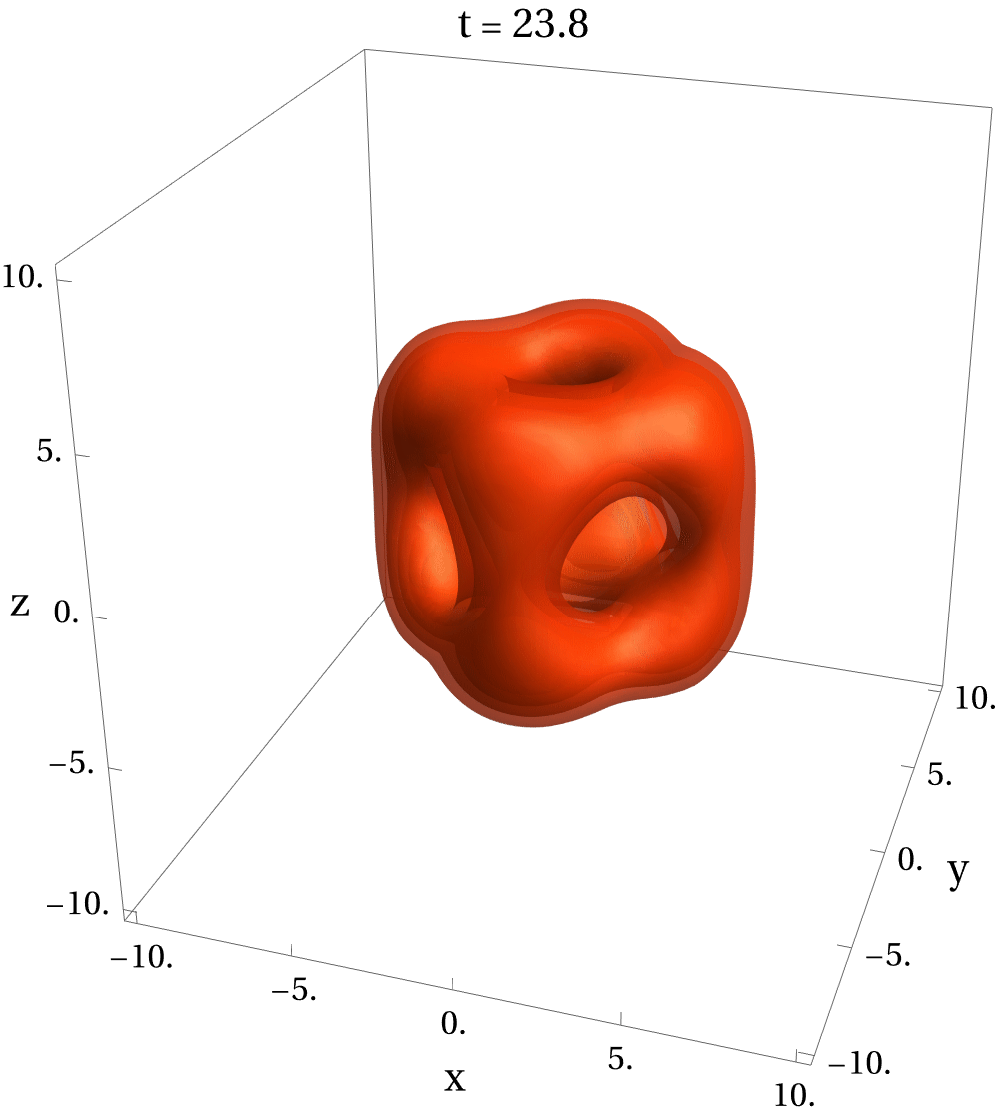}
  \end{subfigure}
  \hspace{\fill}
  \begin{subfigure}{0.3\textwidth}
    \includegraphics[trim=0 0 0 0,clip,width=\textwidth]{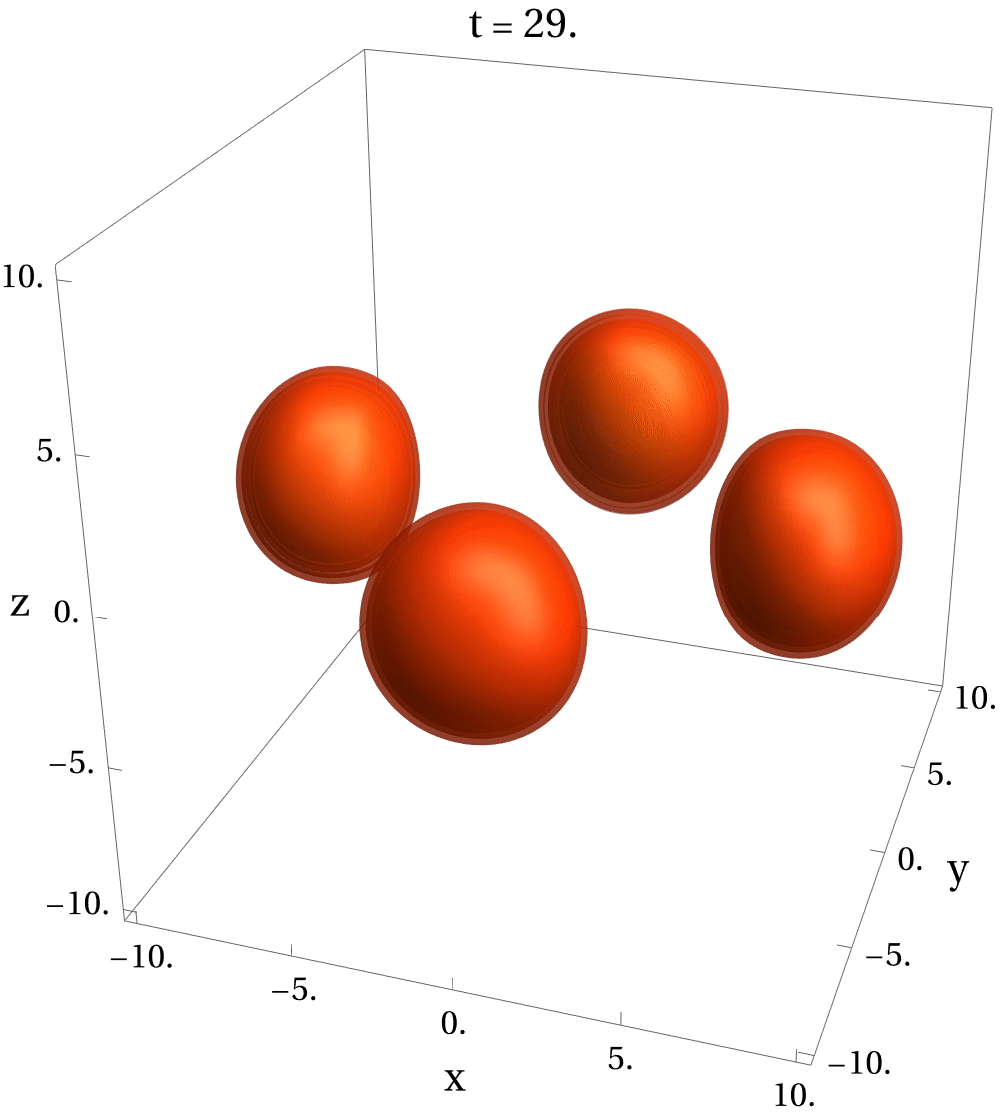}
  \end{subfigure}
  \caption{These contour plots show three frames for the energy densities bigger than $0.014 m_v^4/g^2$ for two charge-$2$ monopoles colliding and splitting into four separated charge-one monopoles. During this process, the energy density goes through a static with cubic energy density shape. The length and time values are given in units of $m_v^{-1}$.}
  \label{fig:22monopole-scattering}
\end{figure*}

In the simulations, we initialized the configurations as described and boosted the monopoles toward each other with a velocity of $0.2$. The expected behavior of the energy density, as outlined above, was observed.
\begin{itemize}
    \item A single monopole colliding with a toroidal charge-two monopole forms a tetrahedron and scatters into three separated monopoles (see Figure~\ref{fig:21monopole-scattering}).
    \item A single monopole colliding with a toroidal charge-three monopole forms a pyramid and scatters into four separated monopoles.
    \item Two toroidal charge-two monopoles form a cube and scatter into four separated monopoles (see Figure~\ref{fig:22monopole-scattering}).
\end{itemize}
Animations corresponding to these three cases can be found in the following video:\\
\url{https://youtu.be/a2wOOo2AAHI} .

From the simulation, we can analyze the behavior of the zeros of the scalar field $\phi^a$, as previously done for planar scatterings. Each zero can be assigned a winding number. As shown in~\cite{Sutcliffe:1996he}, some non-planar scattering scenarios require the introduction of anti-zeros -- zeros with negative winding numbers -- in order to preserve the total number of zeros.

In the three-monopole planar scattering, the number of zeros remained three, except at the point where the monopoles formed an exact torus. At that moment, all zeros coincided at the origin, forming a single zero with winding number three. In contrast, in non-planar scattering, we may also find anti-zeros. 

In the non-planar three-monopole scattering, a charge-two monopole collides with a single monopole. 
Initially, the configuration consists of a zero at the single monopole and a zero of winding-two located at the center of the charge-two monopole. As the distance between the two objects decreases, all the zeros shift from their initial positions. The zero associated with the single monopole moves away from its monopole core toward the charge-two monopole. Meanwhile, the winding-two zero in the torus splits into three zeros of winding one, which arrange themselves cyclically around a single zero of negative winding. When the configuration forms the tetrahedral state, the distances from the anti-zero in the center to the zeros are all the same. Subsequently, the zero originating from the single monopole annihilates with the central anti-zero. This leaves three separated zeros of winding-one, corresponding to three single monopoles that move apart in the $x$-$y$-plane.

In the non-planar four-monopole scattering, we discussed two possible scenarios. In the collision of a single monopole with a toroidal charge-three monopole, the configuration transitions through a pyramidical state, which then splits into four single monopoles.
The behavior of the zeros closely resembles that of the tetrahedral scattering. Initially, the zero of winding-three at the center of the torus splits into four zeros of winding-one and a single anti-zero. The anti-zero then annihilates with the zero originating from the single monopole. At the end, four zeros corresponding to the four monopoles remain and move apart within a single plane.

The other scenario is cubic scattering, in which two toroidal charge-two monopoles form a cube during the interaction and then split into four single monopoles. At the moment of cube-formation, all zeros meet at the center. Subsequently, they split while maintaining cyclic symmetry throughout the process. It is worth mentioning that no anti-zero occurs at any stage of this scattering.

All these observations about the behavior of the zeros agree with the predictions of Paul Sutcliffe in~\cite{Sutcliffe:1996he}.

\section{\MakeUppercase{Conclusions and outlook}}
\label{sec:outlook-and-conclusion}
Using numerical simulations, we validated several predictions of the moduli approximation for multi-monopole collisions. We observed $90^\circ$, $60^\circ$, and $45^\circ$ scattering for two, three, and four monopoles in head-on collisions. Additionally, we simulated tetrahedral and cubic scattering for three and four monopoles, respectively. Although our discussion focuses on purely magnetic monopoles, the formation of tori, tetrahedra, and cubes occurs similarly for skyrmions~\cite{Houghton:1997kg}.

Our simulations offered the advantage of going beyond the moduli approximation, allowing us to analyze these phenomena at relativistic velocities and with non-BPS parameters. Despite the expected magnetic repulsion between two equally charged monopoles for $m_h \neq 0$, we observed similar behaviors in the scattering processes. 


Our analysis validates the robustness of the moduli approximation for monopole collisions. Furthermore, the analytic initial configurations presented in this paper represent a significant step towards a full understanding of multi-monopole systems at the level of the field. While non-planar scattering was achieved using initial setups involving tori and single monopoles, finding an analytic initial configuration for multiple single-charged monopoles undergoing non-planar scattering remains an open challenge.

Additionally, the outcomes of non-head-on collisions remain to be investigated. While an offset within the $x$-$y$-plane results in varied scattering angles, which we were also able to observe in our simulations, the moduli approximation predicts right-angle scattering for an initial offset in the $x$-$z$-plane. In this scenario, the angular momentum transforms into an electric charge on the monopoles, i.e.~the magnetic monopoles become dyons~\cite{Gibbons:1986df, Atiyah-Hitchin1988}.

Recently, it has been shown that the motion of vortices described by the geodesics in the moduli space undergoes substantial modification if massive normal modes of the vortices are excited. Specifically, the characteristic right-angle scattering is replaced by multi-bounce windows that form a chaotic pattern~\cite{AlonsoIzquierdo:2024nbn, Krusch:2024vuy}.
Magnetic monopoles don't possess massive normal modes. However, they can have long-lived semi-bound excitations. In~\cite{GarciaMartin-Caro:2025zkc}, it is predicted that such excitations may lead to changes in the trajectories predicted by the moduli space approximation similar to the vortex scenario. With the configurations presented in this paper, it may be possible to check this behavior numerically in the near future.

\section*{\MakeUppercase{Acknowledgements}}
We want to thank Nick Manton for his invaluable comments, insightful ideas, and helpful suggestions, which greatly contributed to the development of this work.

This work was supported in part by the Humboldt Foundation under Humboldt Professorship Award, 
by the European Research Council Gravities Horizon Grant AO number: 850 173-6,  by the Deutsche Forschungsgemeinschaft (DFG, German Research Foundation) under Germany's Excellence Strategy - EXC-2111 - 390814868, and Germany's Excellence Strategy under Excellence Cluster Origins.

This work is also supported in part by the Israel Science Foundation (grant no. 2159/22), by the Minerva Foundation, by a German-Israeli Project Cooperation (DIP) grant "Holography and the Swampland", and by a research grant from the Chaim Mida Prize in Theoretical Physics at the Weizmann Institute of Science.

J.S.V.B. acknowledges the support from the Departament de Recerca i Universitats from Generalitat de Catalunya to the Grup de Recerca ‘Grup de Fisica Teorica UAB/IFAE’ (Codi: 2021 SGR 00649) and from grant PID2023-146686NB-C31, funded by MICIU/AEI/10.13039/501100011033/ and by FEDER, UE. IFAE is partially funded by the CERCA program of the Generalitat de Catalunya. This study was supported by MICIIN with funding from European Union NextGenerationEU(PRTR-C17.I1) and by Generalitat de Catalunya
\\[0.2cm]
\noindent\textbf{Disclaimer:}
Funded by the European Union. Views and opinions expressed are however those of the authors only and do not necessarily reflect those of the European Union or European Research Council. Neither the European Union nor the granting authority can be held responsible for them.

\section*{\MakeUppercase{Appendix: Numerical Methods}}
\label{sec:numerical-methods}
\textit{Numerical relaxation method.}
Using an iterative relaxation method, we obtained static solutions for charge-$n$ monopoles, as described in section~\ref{sec:charge-n-magnetic-monopoles}. This approach, adapted from~\cite{Saurabh:2017ryg, Patel:2023sfm}, proceeds as follows. The static field equations are first written in the form
\begin{align} 
    \label{eq:relaxation_field_equation} 
    E[f] = \partial_i^2 f + S[f] = 0, 
\end{align} 
where $f$ denotes the field to be relaxed, and $S[f]$ contains all remaining terms. In the model studied here, we choose $f$ to be $\phi^a$ and $W_\mu^a$.
We can write the Laplacian as
\begin{align*}
    \partial_i^2 f=-\frac{6}{\delta^2}f(x,y,z)+\frac{\Delta}{\delta^2},
\end{align*}
with
\begin{align*}
    \Delta=&f(x+\delta,y,z)+f(x-\delta,y,z)+f(x,y+\delta,z)\\
    +&f(x,y-\delta,z)+f(x,y,z+\delta)+f(x,y,z-\delta)
\end{align*}
Inserting this into \eqref{eq:relaxation_field_equation} and solving for $f$ gives
\begin{align*}
    f&=\frac{\Delta}{6}+\frac{\delta^2}{6}S[f]=\frac{\delta^2}{6}E[f]+f.
\end{align*}
The equation is trivially satisfied if $f$ is a solution of the field equations. However, starting with a configuration that does not solve the field equations means $E[f]$ is non-zero, serving as a measure of how well $f$ satisfies the equation. We apply an iterative procedure:
\begin{align} 
    f^{(n+1)} = \frac{\delta^2}{6} E[f^{(n)}] +f^{(n)}, 
\end{align}
which updates $f^{(n)}$ by a correction proportional to the error. Once $f^{(n)}$ satisfies $E[f^{(n)}] = 0$, the solution is found.

In practice, we perform a finite number of iterations. For the solutions presented in section~\ref{sec:charge-n-magnetic-monopoles}, we used $1000$ iterations, with a lattice spacing of $\delta = 0.25 m_v^{-1}$. We observed that more iterations or smaller lattice spacings lead to small improvements to the precision. However, for the scope of this work, the chosen precision is sufficient.
The lattice was a cubic box with $x, y, z \in [-30 m_v^{-1}, 30 m_v^{-1}]$.

In the BPS case, relaxation was applied to the fields $\phi^a$ and $W_i^a$. In contrast, in the non-BPS case, the charge $n$ monopole solutions with $n\geq 2$ are unstable. Relaxing all the fields in these cases causes the higher charge monopole to split into $n$ separate charge-one monopoles. To prevent this splitting, we fix the direction of the scalar field $\hat{\phi}^a$ and apply relaxation only on $\abs{\phi}=\sqrt{\phi^a \phi^a}$.
Expressing $\phi^a=\abs{\phi}\hat{\phi}^a$ in equation~\eqref{eq:field-equations}, we derive the required equation for $\abs{\phi}$. Throughout this procedure, $\hat{\phi}^a$, $\partial_i \hat{\phi}^a$, and $\partial_i^2 \hat{\phi}^a$ are held fixed.\\

\textit{Numerical simulation.}
For numerical time integration, we used the iterated Crank-Nicholson method with two iterations, as described in~\cite{Teukolsky:1999rm}. This method is applied to field equations of the form
\begin{align} 
    \partial_t^2 f = S[f] 
\end{align}
as follows.
First, we rewrite this second-order differential equation as a system of two first-order differential equations
\begin{align*}
    \partial_t \dot{f}&=S[f],\\
    \partial_t f&=\dot{f}.
\end{align*}
We start the iteration with 
\begin{align*}
    \Tilde{\dot{f}}^{(1)}&=\dd t\, S[f_0]+\dot{f}_0,\\
    \Tilde{f}^{(1)}&=\dd t\, \dot{f}_0+f_0, 
\end{align*}
where $f_0$ is the initial field. Next, we take the average of $f_0$ and $\Tilde{f}_0$:
\begin{align*}
    \Bar{\dot{f}}^{(1)}=\frac{\Tilde{\dot{f}}^{(1)}+\dot{f}_0}{2},&& \Bar{f}^{(1)}=\frac{\Tilde{f}^{(1)}+f_0}{2}.
\end{align*}
For the $n$th iteration step, we repeat the same procedure, but in the field equations, $f_0$ is replaced by $\Bar{f}^{(n-1)}$:
\begin{align*}
    \Tilde{\dot{f}}^{(n)}&=\dd t\, S[\Bar{f}^{(n-1)}]+\dot{f}_0,\\
    \Tilde{f}^{(n)}&=\dd t\, \Bar{\dot{f}}^{(n-1)}+f_0, \\
\Bar{\dot{f}}^{(n)}&=\frac{\Tilde{\dot{f}}^{(n)}+\dot{f}_0}{2},&& \Bar{f}^{(n)}=\frac{\Tilde{f}^{(n)}+f_0}{2}.
\end{align*}
The closing step is then given by
\begin{align*}
    \dot{f}_1&=\dd t \, S[\Bar{f}^{(n)}]+\dot{f}_0,\\
    f_1&=\dd t \Bar{\dot{f}}^{(n)}+f_0.
\end{align*}
This iteration procedure is applied at every time step.

As shown in~\cite{Teukolsky:1999rm}, this method is stable for $\dd t < 2 \dd x$ (notice that we use the centered version for discrete spatial derivatives leading to the factor of two). Furthermore, performing more than two iterations does not enhance accuracy. Thus, two iterations are sufficient, and this is what we used for our simulations.

In all simulations presented in this paper, we used a cubic lattice of size $x, y, z \in [-30 m_v^{-1}, 30 m_v^{-1}]$ and a lattice spacing of $0.25 m_v^{-1}$. The time step was set to $\dd t = 0.1 m_v^{-1}$. For the boundary conditions, Dirichlet boundaries were employed.
As a crosscheck, we also tested absorbing boundaries and larger lattices but observed no significant improvements in the results.

To analyze the scattering of monopoles, we apply Lorentz boosts to their initial configurations. This is achieved by modifying the specific coordinates in the initial ansatz. For instance, in the two-monopole configuration described in section~\ref{sec:two-monopole-configuration}, we replace $x - x_{1,2}$ with $\gamma (x - x_{1,2} - u_{1,2}t)$, where $u_{1,2}$ are the velocities of the two monopoles and $\gamma$ is the Lorentz factor.
In the $N$-monopole scenario, $x_n$ in~\eqref{eq:coordinates-n-monopole} is replaced by
\begin{align*}
    x_n = \gamma \big( \cos(\alpha_n)x + \sin(\alpha_n)y - R - ut \big),
\end{align*}
where $u$ is the radial velocity of all monopoles.
Substituting these boosted coordinates into the ansatz provides the initial fields and their time derivatives.

For the gauge condition, we chose the Lorenz gauge $\partial_\mu W^\mu_a = 0$. In order to ensure that this gauge is satisfied initially, we computed $\partial_t W^a_t$ for the initial ansatz using
\begin{align*}
\partial_t W^a_t = \partial_i W^a_i.
\end{align*}

\textit{Programming language.}
The numerical simulations were conducted using the Python programming language, with performance optimizations achieved through the Numba package~\cite{Numba}. Numba accelerates computations by translating the Python code into an efficient machine code, while also providing an easy-to-use framework for parallelizing the code to take full advantage of multi-core processors. Visualization and plotting of the results were carried out using Mathematica.

\setlength{\bibsep}{4pt}
\bibliography{references}

\end{document}